\documentclass[aps,prb,twocolumn,superscriptaddress,longbibliographys]{revtex4-2}
\usepackage{graphicx}
\usepackage{bm}
\usepackage{color}
\usepackage{epstopdf}
\usepackage{amsmath}
\usepackage{url}
\usepackage{braket}
\usepackage{esvect}
\definecolor{darkGreen}{RGB}{0,110,0}
\definecolor{darkBlue}{RGB}{0,0,130}
\usepackage[colorlinks=true, urlcolor=darkBlue,citecolor=darkGreen,linkcolor=darkBlue]{hyperref}
\begin{document}

\title{Spin Hall and Edelstein effects in a ballistic quantum dot with Rashba spin-orbit coupling}


\author{Alfonso Maiellaro}
\affiliation{CNR-SPIN, c/o Università di Salerno, IT-84084 Fisciano (SA), Italy}
\affiliation{Dipartimento di Fisica "E.R. Caianiello", Università di Salerno, Via Giovanni Paolo II, 132, I-84084 Fisciano (SA), Italy}

\author{Francesco Romeo}
\affiliation{Dipartimento di Fisica "E.R. Caianiello", Università di Salerno, Via Giovanni Paolo II, 132, I-84084 Fisciano (SA), Italy}
\affiliation{INFN, Sezione di Napoli, Gruppo collegato di Salerno,Italy}

\author{Mattia Trama}
\affiliation{Dipartimento di Fisica "E.R. Caianiello", Università di Salerno, Via Giovanni Paolo II, 132, I-84084 Fisciano (SA), Italy}

\author{Jacopo Settino}
\affiliation{Dipartimento di Fisica, Università della Calabria, Via P. Bucci Arcavacata di Rende (CS), Italy}

\author{Claudio Guarcello}
\affiliation{Dipartimento di Fisica "E.R. Caianiello", Università di Salerno, Via Giovanni Paolo II, 132, I-84084 Fisciano (SA), Italy}
\affiliation{INFN, Sezione di Napoli, Gruppo collegato di Salerno,Italy}

\author{Carmine Antonio Perroni}
\affiliation{Dipartimento di Fisica Ettore Pancini, Università degli Studi di Napoli Federico II, Via Cinthia, 80126-Napoli, Italy}

\author{Pawel Wójcik}
\affiliation{AGH University of Krakow, Faculty of Physics and Applied Computer Science, Al. Mickiewicza 30, 30-059 Krakow, Poland}

\author{Bartłomiej Szafran}
\affiliation{AGH University of Krakow, Faculty of Physics and Applied Computer Science, Al. Mickiewicza 30, 30-059 Krakow, Poland}

\author{Daniela Stornaiuolo}
\affiliation{Dipartimento di Fisica Ettore Pancini, Università degli Studi di Napoli Federico II, Via Cinthia, 80126-Napoli, Italy}

\author{Marco Salluzzo}
\affiliation{CNR-SPIN, Complesso Monte Sant’Angelo-Via Cinthia, I-80126 Napoli, Italy}

\author{Thomas Sand Jespersen}
\affiliation{Department of Energy Conversion and Storage, Technical University of Denmark, 2800 Kgs. Lyngby, Denmark}

\author{Nicolas Bergeal}
\affiliation{Laboratoire de Physique et d’Etude des Matériaux, ESPCI Paris, Université PSL, CNRS, Sorbonne Université, Paris, France}

\author{Manuel Bibes}
\affiliation{Laboratoire Albert Fert, CNRS, Thales, Université Paris-Saclay, 91767 Palaiseau, France}

\author{Roberta Citro}
\affiliation{Dipartimento di Fisica "E.R. Caianiello", Università di Salerno, Via Giovanni Paolo II, 132, I-84084 Fisciano (SA), Italy}
\affiliation{INFN, Sezione di Napoli, Gruppo collegato di Salerno,Italy}
\date{\today}

\begin{abstract}
We study spin-resolved transport in a ballistic quantum dot with Rashba spin--orbit coupling, focusing on charge-to-spin conversion and spin Hall effect. In the regime where the dot size is comparable to the Fermi wavelength, we identify a clear crossover from weak localization to weak antilocalization as the Rashba coupling increases. This transition is accompanied by gate-tunable spin currents of Edelstein and spin Hall type, whose behavior reflects the underlying electron wavefunction interference. Notably, the Edelstein current shows an inflection point at the critical Rashba strength, signaling the crossover from weak localization to weak antilocalization. In the presence of an in-plane magnetic field we also report a transition in angular periodicity of the magnetoresistance --from $\pi$ to $2\pi$-- arising from the interplay between spin--orbit interaction and Zeeman coupling. These results establish a direct link between quantum coherence, charge-to-spin conversion, and geometric confinement in mesoscopic systems.
\end{abstract}

\maketitle
\section{Introduction}
Quantum coherence plays a central role in low-dimensional systems, where phase-stable electronic trajectories give rise to interference phenomena that strongly affect transport properties~\cite{PhysRevB.46.6846,PhysRevLett.107.027204,Trama,Liang_2009,Nitta2003,Bychkov1984,Awschalom2007}. Among the mechanisms that enrich such phenomena, spin–orbit coupling (SOC)~\cite{Bychkov1984} plays a crucial role, not only in coherent transport but also in a variety of quantum materials—including topological superconductors~\cite{Lutchyn2010,Alicea2012,Yokoyama2009}, topological insulators and phase-coherent Josephson junctions~\cite{SciPostPhysMa,GUARCELLO}.
\begin{figure}[!htbp]
	\centering
	\includegraphics[width=0.45\textwidth]{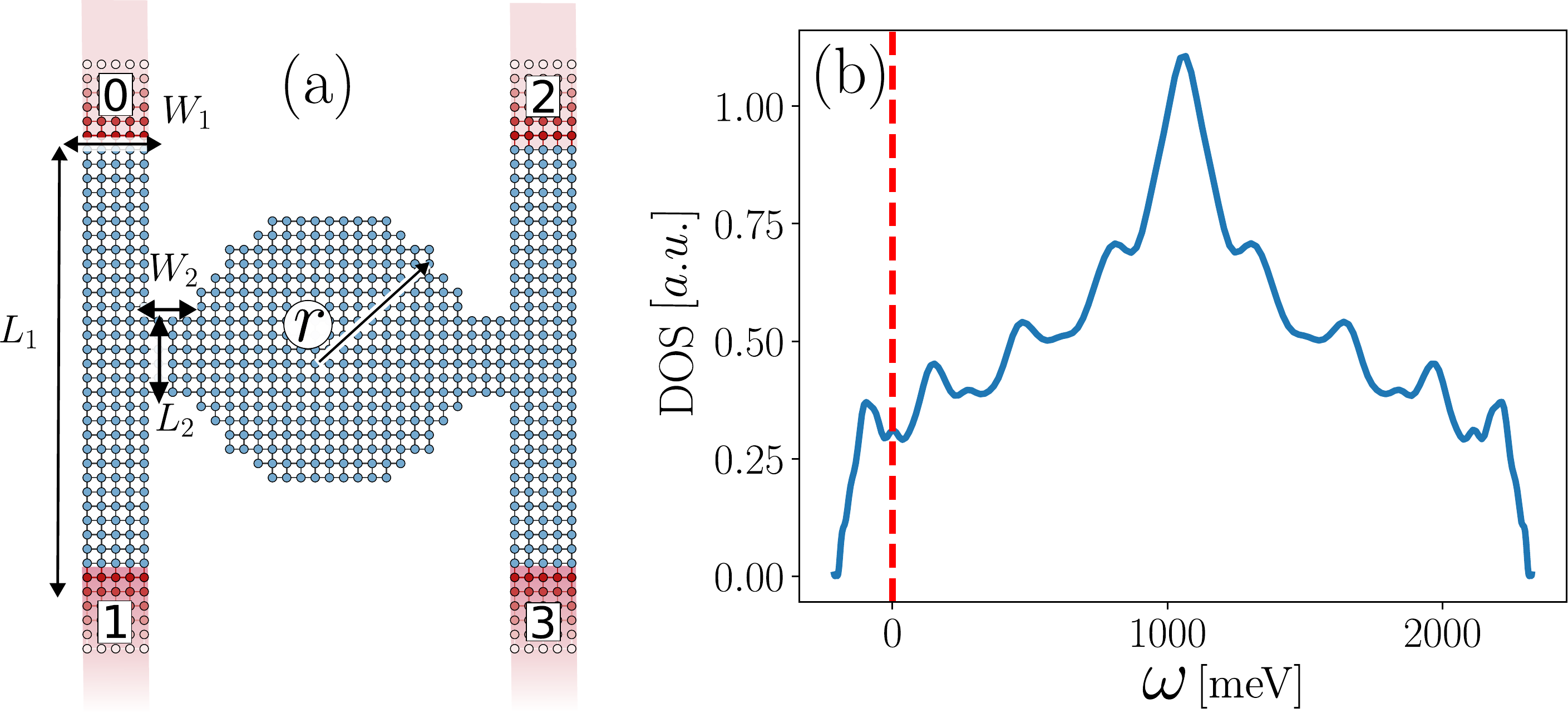}
	\caption{(a) Tight-binding lattice used in our model, implemented via Kwant~\cite{Groth_2014}. The central blue region defines the quantum dot (scattering region), composed of a circular core of radius $r$ connected to two symmetric compound arms, each consisting of two adjacent rectangles. The four red regions indicate the semi-infinite normal-metal leads, labeled from 0 to 3. Geometric parameters $L_1$, $L_2$, $W_1$, and $W_2$ define the arm sizes, as shown. (b) Density of states (DOS) of the dot. The vertical dashed red line marks the chemical potential considered here.}
	\label{Figure0}
\end{figure}
 In systems with SOC, spin and orbital dynamics become entangled: interference between time-reversed trajectories can be modulated by spin precession, and external fields or geometric asymmetries can tune the resulting transport signatures. This complexity is further enhanced when SOC is present in confined geometries, where boundary conditions and finite-size effects introduce additional structure to spin-dependent dynamics~\cite{Li2018,PhysRevLett.87.256801}. In this context, the interplay between quantum coherence, SOC, and confinement gives rise to unconventional spin and charge transport effects~\cite{Maiellaro2025,PhysRevB.98.184418,PhysRevApplied}. The situation is even more interesting since spin-orbit coupling has been proven to be tunable using all-electrical means~\cite{PhysRevLett.90.076807,PhysRevLett.89.276803,nanolett.9b04079}. In particular, spin-to-charge conversion mechanisms—such as Edelstein~\cite{EDELSTEIN1990233} and spin Hall~\cite{Sinova2004} effects—become sensitive to interference conditions and can exhibit nontrivial dependencies on system size, geometry, and Rashba coupling strength. In this context, Ref.~\cite{Maiellaro2025} introduced a theoretical framework based on a spin-dependent scattering-matrix approach to describe charge-to-spin conversion mechanisms, with applications focused on two-dimensional nanostructures.\\
A key manifestation of spin-dependent interference is the crossover from weak localization (WL) to weak antilocalization (WAL)~\cite{PhysRev.115.485,Bergmann1984,Hikami1980}. This crossover is observed, for example, in two-dimensional electron gases (2DEGs) in semiconductor heterostructures such as GaAs/AlGaAs or InAs quantum wells, where gate voltage or carrier density tunes the Rashba spin--orbit coupling (SOC) strength~\cite{PhysRevLett.53.1100,PhysRevLett.89.046801}. It also occurs in thin films of topological insulators (e.g., Bi$_2$Se$_3$, Bi$_2$Te$_3$), as disorder or Fermi level position changes~\cite{PhysRevB.84.233101,PhysRevLett.105.176602}, and in oxide interfaces such as LaAlO$_3$/SrTiO$_3$, where the Rashba coupling can be controlled by gates~\cite{PhysRevLett.104.126803,PhysRevB.97.075136}. The crossover thus provides a powerful probe of spin coherence and SOC symmetry in mesoscopic systems. While traditionally associated with diffusive systems and impurity scattering, studies have shown that similar WL–WAL features can emerge in clean, ballistic conductors—provided that confinement supports phase-coherent backscattering~\cite{Baranger1993,Zozoulenko1996,Cserti2004}.\\
Motivated by these studies and by the need of understanding the confinement effects on WL-WAL transition, we study spin-resolved quantum transport in a ballistic quantum dot (QD) with Rashba SOC, focusing on the regime where the dot size is comparable to the Fermi wavelength $\lambda_F$. Building on the scattering-matrix framework introduced in Ref.~\cite{Maiellaro2025}, here applied to a strongly confined geometry, we find that quantum confinement and spin–orbit interaction conspire to generate gate-tunable interference effects and spin-polarized currents. This allows us to establish a direct connection between spin–charge conversion mechanisms and the WL–WAL crossover in mesoscopic ballistic systems. In particular, we apply the analysis to QDs based on LAO/STO where a 
control of SOC can be fully achieved by back gate~\cite{PhysRevB.90.235426,Rev2}.\\
The paper is organized as follow: In Sec.~\ref{Theory} we present the tight binding Hamiltonian 
and the theoretical framework used to describe the multi-terminal quantum transport in a QD. In Sec.~\ref{NumRes} we analyze the magnetotransport curves and the microscopic spin current patterns. Conclusions are drawn in ~\ref{Conclusions}, while supplemental material is reported in Appendices~\ref{AppA}-~\ref{AppC}.\\
\section{Theoretical model}
\label{Theory}
\begin{figure*}
	\centering
	\includegraphics[width=0.9\textwidth]{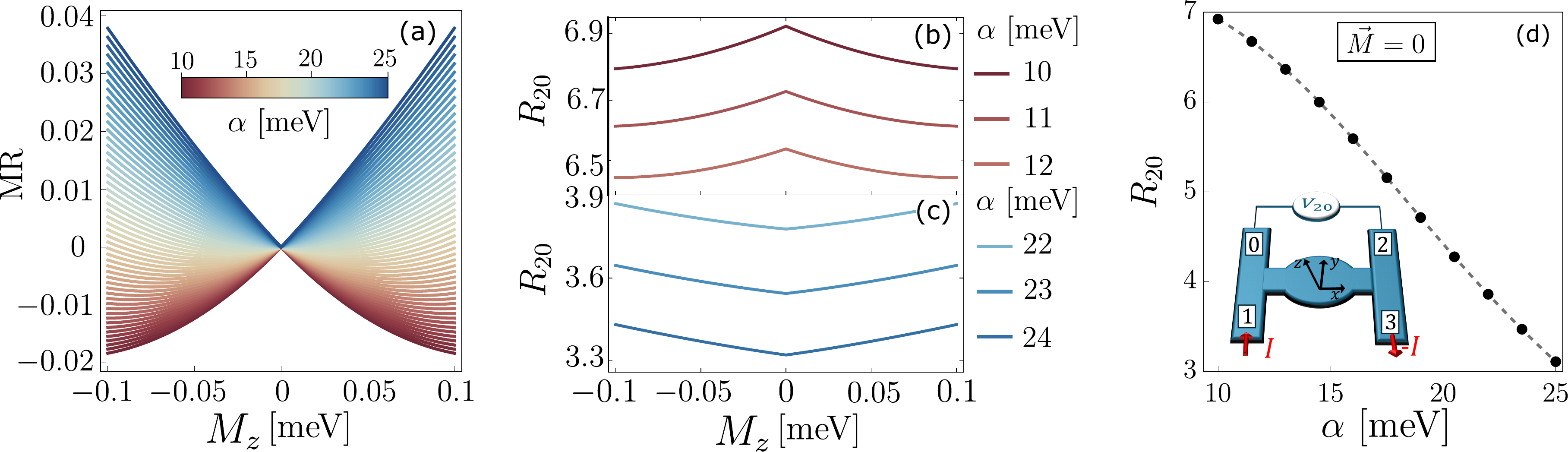}
	\caption{(a) Magnetoresistance MR between leads 2 and 0 as a function of Zeeman energy $M_z$, for $\alpha$ ranging from 10 to 25 meV. The curvature reversal around $M_z = 0$ highlights a crossover from WL to WAL as $\alpha$ increases. This transition is also reflected in the color shift from warm tones (WL) to cool tones (WAL). (b)–(c) Resistance $R_{20}$ vs $M_z$ for selected values of $\alpha$ from panel (a), illustrating the curvature evolution that characterizes the WL-to-WAL transition. (d) Resistance $R_{20}$ as a function of $\alpha$ for $\vec{M} = 0$, showing a monotonic decrease due to the enhanced suppression of backscattering by spin-orbit interaction. The inset illustrates the four-terminal setup and current injection scheme. The resistance curves are expressed in unit of $R_0=h/e^2$.}
	\label{Figure1}
\end{figure*}
We model the QD (represented in Figure~\ref{Figure0}) using a tight-binding Hamiltonian on a square lattice, including kinetic energy, Rashba spin--orbit interaction, and Zeeman coupling. The full Hamiltonian reads $H = H_0 + H_{\text{SO}} + H_M$, where
\begin{eqnarray}
	\label{HamKinetic}
	\begin{split}
		H_0\!=\! &\sum_{x,y} \Bigl[ (-\epsilon+4t) \Psi^{\dagger}_{x,y} \sigma_0 \Psi_{x,y} -t \Bigl(\Psi^{\dagger}_{x,y} \sigma_0 \Psi_{x+1,y} + H.c\Bigr)+\\
		&-t \Bigl(\Psi^{\dagger}_{x,y} \sigma_0 \Psi_{x,y+1}+ H.c.\Bigr) \Bigr] 
	\end{split}
	\label{HamK}
\end{eqnarray}
\begin{eqnarray}
	\label{Rashba}
	H_{SO}\!=\! \mathrm{i} \alpha \Bigl(\!\sum_{x,y}\! \Psi^{\dagger}_{x,y} \sigma_y \Psi_{x+1,y}\!\! -\!\Psi^{\dagger}_{x,y} \sigma_x \Psi_{x,y+1}\!\Bigr)\!\!+\!\! H.c.
	\label{HamSOC}
\end{eqnarray}
\begin{eqnarray}
	\label{Zeeman}
	H_{M}= \sum_{x,y} \Psi^{\dagger}_{x,y}\ \vec{M} \cdot \vec{\sigma}\ \Psi_{x,y}.
	\label{HamZeeman} 
\end{eqnarray}	
Here, the indices $(x,y)$ span the lattice sites of the quantum dot, corresponding to the blue region in Fig.~\ref{Figure0}(a). $\Psi_{x,y} = (c_{x,y,\uparrow}, c_{x,y,\downarrow})^T$ is the spinor of annihilation operators, $\sigma_0$ is the identity matrix, and $\vec{\sigma} = (\sigma_x, \sigma_y, \sigma_z)$ is the vector of Pauli matrices. The parameters $t$, $\alpha$, and $\vv{M} = (M_x, M_y, M_z)$ denote the hopping amplitude, Rashba coupling, and Zeeman field, respectively. The on-site energy is set to $\epsilon = \epsilon_0+\mu$, where $\mu$, determining the filling of the Rashba quantum dot,  represents the energy offset measured from the lowest energy eigenvalue $\epsilon_0$. The parameter $\epsilon_0$ is consistently computed across all realizations of $\alpha$ and $\vec{M}$. When an external magnetic field is considered, the orbital magnetic effects are embedded in the Peierls phase, $t \rightarrow t\, e^{i \frac{e}{\hbar} \int_{\vec{r}_i}^{\vec{r}_j} \vec{A} \cdot d\vec{r}} \equiv t(\vec{r}_i,\vec{r}_j)$, with $\vec{A}$ the vector potential.\\ The dot geometry is modeled as a central circular region of radius $r$ connected to two lateral arms of dimensions $(L_1, W_1)$ and $(L_2, W_2)$. This setup incorporates the realistic electrostatic confinement obtained in the experiments~\cite{Prawiroatmodjo2017} and includes the surface roughness by randomly breaking or preserving hopping terms at the boundary of the circular region (see Fig.~\ref{Figure0}). In this geometry, the effect of the confinement is reflected in the formation of subbands, inducing peaks in the density of states of the QD as shown in Figure 1(b). 
Hereafter we refer to system parameters relevant for oxide interfaces, but results remain valid also for other systems. In particular,  we use a tight-binding hopping parameter: $t \approx 313$ meV, based on an effective mass $m_{\text{eff}} = 0.8\, m_e$ and lattice constant $a = 0.39$\,nm~\cite{PhysRevB.103.235120}. Unless otherwise specified, we also assume $\mu=180$ meV, $r/a = 10$, $L_1/a = 30$, $W_1/a = 5$, $L_2/a = 6$, and $W_2/a = 3$ and explore a broad range of $\alpha$, which are experimentally tunable via gate voltage~\cite{PhysRevLett.90.076807,PhysRevLett.89.276803,nanolett.9b04079}. The values of $\alpha$ considered here are consistent with those reported for LAO/STO interfaces~\cite{Jouan2020,PhysRevB.103.235120} taking into account that confinement effects can enhance the effective Rashba coupling. These values are also comparable to those observed in KTO-based interfaces, where even larger Rashba couplings have been reported~\cite{Arche}. We focus on the regime where the dot size is comparable to the Fermi wavelength, which enhances quantum interference effects and makes the WL–WAL crossover particularly evident. While larger QD sizes have been typically considered for LAO/STO, the present choice enables an efficient numerical treatment and captures the relevant physics governed by a ballistic regime\\
The QD is connected to four semi-infinite nonmagnetic metal leads, as shown in Fig.~\ref{Figure0}(a). When a dc voltage $V^j$ is applied to each lead $j = 0,\dots,3$, the charge current $\langle J^j_c \rangle$, spin current $\langle \vec{J}^j_s \rangle$, and bias-induced spin density $\langle \vec{\delta s}^{j} \rangle$ in lead $j$ can be computed within the scattering framework of Ref.~\cite{Maiellaro2025} as:
\begin{eqnarray}
	\langle J^j_c \rangle &=& \sum_{j'} \frac{e^2}{2 \pi \hbar}\!\! \left[ 2 \mathcal{N}^j \delta^{jj'}\!\! -\!\! \sum_{m,m'} \text{Tr}(S_{mm'}^{jj'\dagger} S_{mm'}^{jj'}) \right]\!\!V^{j'}\!\!, \label{Equation1} \\
	\langle \vec{J}^j_s \rangle &=& \sum_{j',m,m'} \frac{e}{4\pi} \text{Tr}(S_{mm'}^{jj'\dagger} \vec{\sigma} S_{mm'}^{jj'}) V^{j'}, \label{Equation2} \\
	\langle \vec{\delta s}^{j} \rangle &=& \sum_{j',m,m'} \frac{e}{4\pi |v^j_m(\mu)|} \text{Tr}(S_{mm'}^{jj'\dagger} \vec{\sigma} S_{mm'}^{jj'}) V^{j'}. \label{Equation3}
\end{eqnarray}
Here, $e$ is the electron charge, $2\mathcal{N}^j$ is the total number of quantum channels in lead $j$, and $v^j_m(\mu)$ is the group velocity of mode $m$ at chemical potential $\mu$. The scattering matrix $S$ is decomposed as $S = \sum_{jj', mm', \sigma, \sigma'} \mathcal{P}_{jj'} \otimes \mathcal{P}_{mm'} \otimes \mathcal{P}_{\sigma \sigma'}\, S^{jj'}_{mm'\sigma\sigma'}$, where $\mathcal{P}_{\eta \eta'} = \ket{\eta} \bra{\eta'}$ is a projection operator in the leads, channel and spin subspaces, respectively. In Eqs.~(\ref{Equation1})--(\ref{Equation3}), the physical observables evaluated at lead $j$  depend on all incident channels $m'$ across the lead $j'$. The low-filling regime is considered, where only a few transverse modes are active—highlighted by the vertical dashed red line in Fig.~\ref{Figure0}(b).\\
The expressions above are derived in the asymptotic region of the lead, far from the scattering center, 
under the assumption of zero temperature and linear response in the applied voltages. The scattering matrix $S$ entering our equations is computed numerically using the Kwant package~\cite{Groth_2014}, which is employed as an efficient toolbox to define the system geometry and extract the corresponding scattering amplitudes.\\ 
In our simulations, we inject a current $I$ from lead 1 to lead 3, such that $\langle J^j_c \rangle = (\delta^{1j} - \delta^{3j}) I$.  The voltages at all terminals are then obtained self-consistently by inverting Eq.~(\ref{Equation1}), while the corresponding spin currents and spin densities are computed from Eqs.~(\ref{Equation2})--(\ref{Equation3}).
\section{Numerical results}
\label{NumRes}
In Fig.~\ref{Figure1}(a), we analyze the magnetoresistance between leads $0$ and $2$, defined as
\[
MR = \frac{R_{20}(\vec{M}) - R_{20}(0)}{R_{20}(0)},
\]
\begin{figure*}
	\centering
	\includegraphics[width=0.95\textwidth]{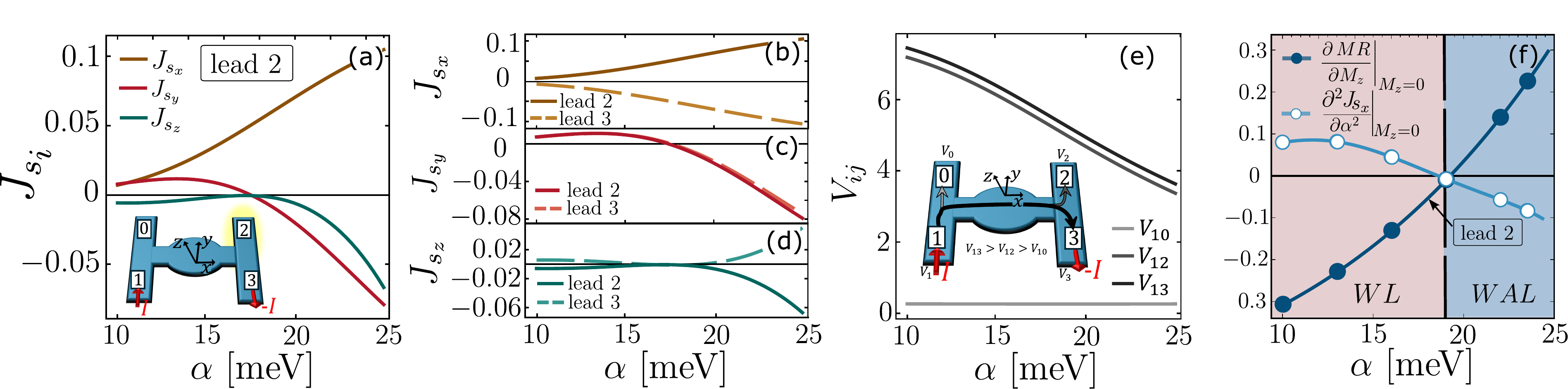}
	\caption{(a) $x$-, $y$-, and $z$-polarized components of the spin current in lead 2 as a function of $\alpha$ for $\vec{M} = 0$. (b)–(d) Comparison of the spin currents in leads 2 (solid lines) and 3 (dashed lines) as a function of $\alpha$, showing how spin-polarized transport evolves between the two leads. 
		(e) Voltage differences between the current-injecting lead (lead 1) and all other leads, illustrating how the internal potential landscape evolves with $\alpha$ and reflects the redistribution of current paths within the scattering region. (f) Comparison between the derivative of the magnetoresistance, $\partial$MR$/\partial M_z$, and the second derivative of the Edelstein spin current, $\partial^2 J_{s_x}/\partial \alpha^2$, both evaluated at $M_z = 0$ and plotted as a function of $\alpha$. The sign change in both quantities identifies the crossover from WL to WAL. In the plot, $\partial^2 J_{s_x}/\partial \alpha^2$ is shown scaled  by 100 for readability. The voltage and spin current curves are expressed in units of $V_0=h I/e^2$ and $J^0_s=e V_0/4 \pi$, respectively, where $I$ is the applied current bias.}
	\label{Figure2}
\end{figure*}
under an out-of-plane Zeeman field $\vec{M} = (0,0,M_z)$. Here $R_{20} = V_{20}/I$, with $V_{20}$ the voltage difference between leads $2$ and $0$. The color scale reveals a clear crossover from WL to WAL as $\alpha$ increases. In the WL regime (small $\alpha$), the resistance $R_{20}$ shows a maximum at $M_z = 0$ and decreases with increasing $|M_z|$, as shown in Fig.~\ref{Figure1}(b). Conversely, in the WAL regime (large $\alpha$), $R_{20}$ exhibits a minimum at $M_z = 0$ and increases with $|M_z|$, as seen in Fig.~\ref{Figure1}(c). This inversion signals the crossover and allows us to identify a critical threshold $\alpha_c \approx 19$~meV for the WL--WAL transition. Figure~\ref{Figure1}(d) further supports this interpretation: the zero-field resistance $R_{20}(M_z=0)$ decreases monotonically with $\alpha$, reflecting the destructive interference associated with WAL. The WL-WAL crossover is governed by the spin–orbit length which can be expressed as $\lambda_{\mathrm{SO}}/a = t/\alpha \in [31.3,12.5]$, where $a$ is the lattice spacing and the precise transition condition depends strongly on the dot size. The WL regime corresponds to $\lambda_{\mathrm{SO}}$ larger than the dot size, where spins remain nearly aligned and interference is constructive, while in the WAL regime $\lambda_{\mathrm{SO}}$ becomes shorter, leading to spin precession and destructive interference. The critical value $\alpha_c \approx 19$~meV thus corresponds to $\lambda_{\mathrm{SO}} \approx 2r$, marking the onset of strong spin–orbit coupling regime within the dot.\\
While such WL--WAL crossovers are typically observed in disordered diffusive systems due to impurity scattering~\cite{PhysRev.115.485,Bergmann1984,Hikami1980}, here, on the other hand, we show that they also occur in a ballistic quantum dot. This happens in a kinematic regime in which the Fermi wavelength  $\lambda_F=2 \pi/k_F$ is comparable to the dot radius $r$, inducing a sequence of scattering processes defining closed clockwise and counterclockwise paths that interfere constructively\footnote{In our simulations, the leads are modeled as semi-infinite electrodes, translationally invariant along the \(y\)-direction, with the transverse motion (along \(x\)) quantized into five propagating modes. Each mode has its own Fermi wavevector \(k_F\), and for the five modes we find \(k_Fa \in [0.7, 1.8]\), which corresponds to \(\lambda_F/a \in [3.48, 8.97]\)} (see Appendix~\ref{AppA}). Such a condition is experimentally relevant for oxide-based quantum dots realized at LAO/STO (001) interfaces~\cite{Prawiroatmodjo2017}, where the dot size is of the same order as the mean free path. We further confirm that increasing the radius $r$ (for $\alpha=0$) suppresses the WL signature, consistent with a transition to a genuine ballistic regime where $\lambda_F \ll r$ (see Appendix~\ref{AppA}). However, if one introduces scattering by impurities or considers a non-uniform background potential in the QD region, forming puddles, the transition can be obtained also in the case of larger sizes. Let us stress that the situation described here is different from the one described by random scattering matrix approach~\cite{RevModPhys.69.731,PhysRevLett.87.256801}. \\
\begin{figure*}[!htbp]
	\centering
	\includegraphics[width=0.9\textwidth]{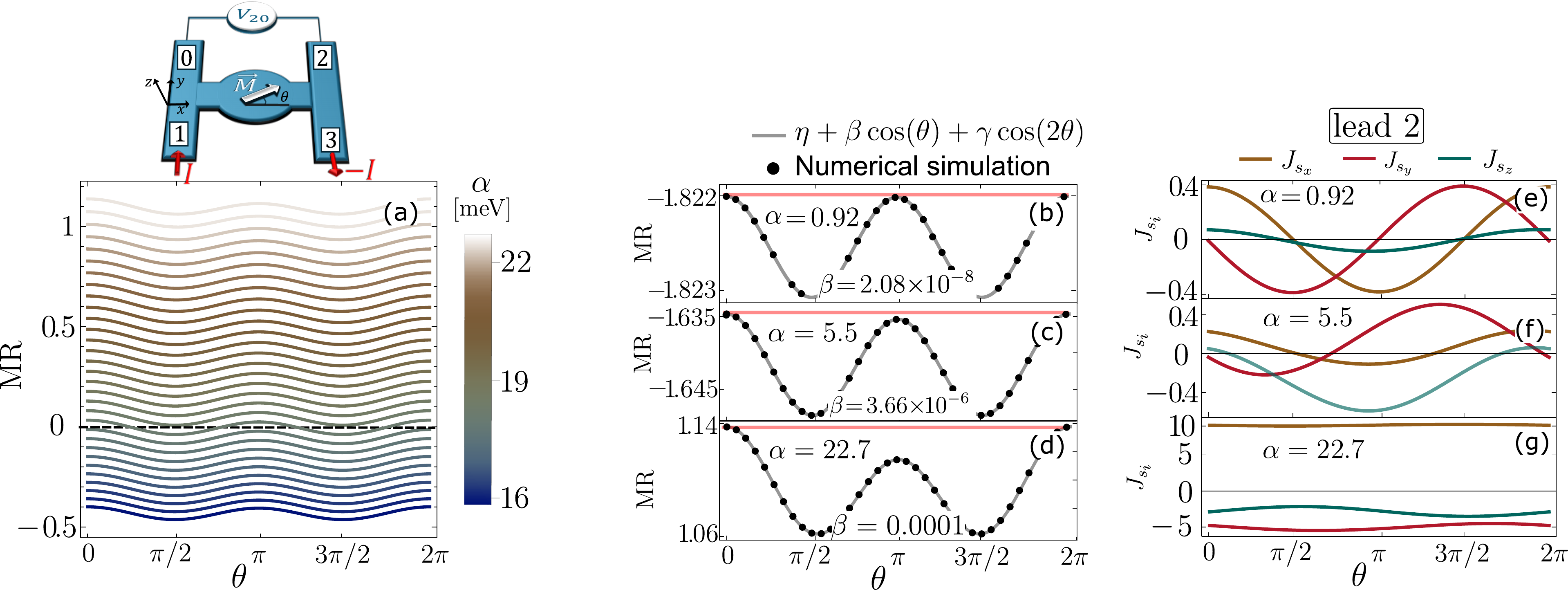}
	\caption{(a) Magnetoresistance MR as a function of the direction $\theta$ of an in-plane Zeeman field $\vec{M} = (M \cos \theta, M \sin \theta, 0)$, for $\alpha$ ranging from 16 to 22 meV. The field amplitude is kept constant at $M = 0.05\ \mathrm{meV}$.
		(b)–(d) MR for three selected values of $\alpha$, showing a transition in angular periodicity from $\pi$ to $2\pi$ as $\alpha$ increases. This behavior reflects the growing dominance of Rashba spin-orbit coupling over the external Zeeman field. The numerical data are well described by a fitting function including $\cos(\theta)$ and $\cos(2\theta)$ harmonics.
		(e)–(g) $x$-, $y$-, and $z$-polarized components of the spin current in lead 2, calculated for the same parameter sets as in panels (b)–(d). At low $\alpha$, the spin currents exhibit harmonic oscillations induced by the in-plane field, particularly in $J_{s_x}$ and $J_{s_z}$, while $J_{s_y}$ remains negligible. At larger $\alpha$, the spin currents become nearly constant with $\theta$, indicating a regime dominated by the spin-orbit interaction.  The spin current curves are expressed in units of $J^0_s=e V_0/4 \pi$, with $V_0=h I/e^2$ and $I$ is the applied current bias. For readability, all y-axis values are scaled by $10^2$.}
	\label{Figure3}
\end{figure*}
Alongside the MR response, in Fig.~\ref{Figure2}(a) we investigate the spin current components $J_{s_i}$ $(i = x,$ $y, z)$ flowing into lead 2 as a function of $\alpha$. The $x$-component, $J_{s_x}$, is due to the Edelstein effect (EE)~\cite{EDELSTEIN1990233}, and is governed by the transverse electric field along the $y$-direction, i.e., the direction of charge current injection. Interestingly, we also detect a finite $J_{s_y}$ component, which is not expected in bulk systems. This contribution is due to a confinement effect, which usually modifies the potential energy landscape so that the carriers cannot escape. This confinement typically results from an electrostatic scalar potential $V(r)$, which creates an electric field pointing towards the center of the confinement or towards boundaries even in the absence of an out-of-plane magnetic field. This electric field locally tilts the spin orientation and gives rise to a nonzero $J_{s_y}$, an effect consistent with predictions from Ref.~\cite{Maiellaro2025}. This effect is a consequence of quantum confinement. Indeed, we have verified that by changing the bias configuration, the spurious in-plane spin current can be fully suppressed, leaving only the Edelstein and spin Hall currents (see Appendix~\ref{AppB}). In addition, a finite out-of-plane spin current $J_{s_z}$ appears as a hallmark of spin Hall physics. In this regime, spin-up and spin-down carriers are deflected in opposite directions, generating a transverse spin Hall current $J_{s_z}$, observed in Fig.~\ref{Figure2}(a). The $\alpha$-dependence of the spin current components is further analyzed in Figs.~\ref{Figure2}(b)--(d), where we compare the signals measured at leads 2 and 3. Both $J_{s_x}$ and $J_{s_z}$ change sign between the two leads, consistently with their origin from EE and SHE, respectively. Conversely, $J_{s_y}$ does not change sign under lead exchange, confirming its confinement-induced nature. To support this interpretation, we analyze in Fig.~\ref{Figure2}(e) the external bias $V_{ij}$ measured between lead pairs in the absence of $\vec{M}$. These potentials qualitatively mimic the internal field profile induced by the current bias $\langle J^j_c \rangle = (\delta^{1j} - \delta^{3j}) I$, suggesting that the sign reversal of $J_{s_x}$ and $J_{s_z}$ originates from a reversal of the internal flux between the two lateral arms. 
A clear fingerprint of the WL–WAL crossover also emerges from a closer inspection of the Edelstein spin current shown in Fig.~\ref{Figure2}(f). In particular, $J_{s_x}$ displays an inflection point at the critical Rashba strength $\alpha_c \approx 19~\mathrm{meV}$, corresponding to the condition $\lambda_{\mathrm{SO}} \approx 2r$, where electrons complete one full spin precession while traversing the dot. A further increase in the spin--orbit coupling shortens $\lambda_{\mathrm{SO}}$, enhancing spin precession and promoting dephasing among different propagation directions. Accordingly, the WL--WAL transition and the Edelstein response share the same microscopic origin, both stemming from the progressive increase of spin precession induced by the Rashba field. Both the inflection point in the spin current and the sign change in the derivative of the magnetoresistance define the boundary between the weak SOC and strong SOC regime.\\
Beyond its impact on spin coherence and out-of-plane spin transport, quantum confinement also influences magnetic anisotropy. This becomes evident under in-plane magnetic fields, where spin textures and SOC interplay give rise to nontrivial angular dependence of the magnetoresistance~\cite{PhysRevB.81.134520,PhysRevB.103.014520,PhysRevB.93.245105}. Motivated by this observation, in Fig.~\ref{Figure3}(a) we study MR as a function of the direction $\theta$ of an in-plane Zeeman field $\vec{M} = M(\cos \theta, \sin \theta, 0)$. We observe that the angular dependence of MR undergoes a periodicity change—from $\pi$ to $2\pi$—as $\alpha$ increases (Figs.~\ref{Figure3}(b)-(d)). This transition reflects the growing influence of spin-orbit interaction over the Zeeman term. The latter observation can be explained by a symmetry argument. Indeed in the absence of Rashba interaction, the Hamiltonians with $\theta=0$ and $\theta=\pi$ are related by a unitary transformation, and this relation is broken in the presence of Rashba SOC (see Appendix \ref{AppC}). This argument explains the periodicity change exhibited by the magnetoresistence curves in Fig. \ref{Figure3}. The associated spin currents also exhibit a crossover: from harmonic, field-driven oscillations with a negligible $z$-component (Fig.\ref{Figure3}(e)), to nearly flat angular profiles with a sizable (confinement-induced) out-of-plane component (Figs.\ref{Figure3}(f)-(g)), signaling the suppression of field-driven spin precession in the Rashba-dominated regime. These results can be understood in terms of expectation values of the spin operators evaluated using the eigenstate of the translational-invariant Rashba Hamiltonian (see Appendix \ref{AppC}), thus confirming this interpretation.
\section{Conclusions}
\label{Conclusions}
We have investigated spin-dependent quantum transport in a ballistic Rashba quantum dot, focusing on quantum interference effects and spin-charge conversion phenomena. When the dot radius is comparable to the Fermi wavelength, a clear WL–WAL crossover emerges as the Rashba coupling strength is increased—a parameter that can be experimentally tuned via gate voltage. In this regime, coherent backscattering paths undergo constructive or destructive interference depending on $\alpha$, leading to a tunable magnetoresistance response. As the dot radius increases, this interference regime is suppressed, unless effects of disorder or puddles formation are considered, and the system enters a fully ballistic transport regime where neither WL nor WAL signatures are observed.\\
This crossover is accompanied by the emergence of spin-polarized currents driven by the Rashba interaction, including both Edelstein and spin Hall contributions. Remarkably, the Edelstein current displays an inflection point at the same critical Rashba strength marking the WL–WAL transition, highlighting a direct connection between spin-charge conversion and quantum interference.\\
Further insight into the role of spin-orbit interaction is provided by the behavior of magnetoresistance curves under in-plane magnetic fields, which exhibit a transition in angular periodicity—from \(\pi\) to \(2\pi\)—as $\alpha$ increases. This behavior is reflected in the evolution of spin current components and offers an effective probe of magnetic anisotropy linked to intrinsic spin textures.\\
Together, our results demonstrate how spin-orbit interaction, quantum coherence, and geometric confinement combine to shape spin and charge transport in mesoscopic systems, with direct relevance for spin-dependent transport in mesoscopic devices.
\section{Acknowledgments.}
The authors acknowledge support from Horizon Europe EIC Pathfinder under the grant IQARO number 101115190. R.C. and F.R. acknowledge funding from Ministero dell’Istruzione, dell’Università e della Ricerca (MIUR) for the PRIN project STIMO (GrantNo. PRIN 2022TWZ9NR). This work received funds from the PNRR MUR project PE0000023-NQSTI (TOPQIN and SPUNTO) and from the project INNOVATOR (National centre HPC, big data and quantum computing). C.A.P. acknowledges funding from the PRIN 2022 PNRR project P2022SB73K “Superconductivity in KTaO3 Oxide-2DEG NAnodevices for Topological quantum Applications” (SONATA) financed by the European Union - Next Generation EU.

\appendix
\begin{figure*}[!htbp]
	\centering
	\includegraphics[width=0.9\textwidth]{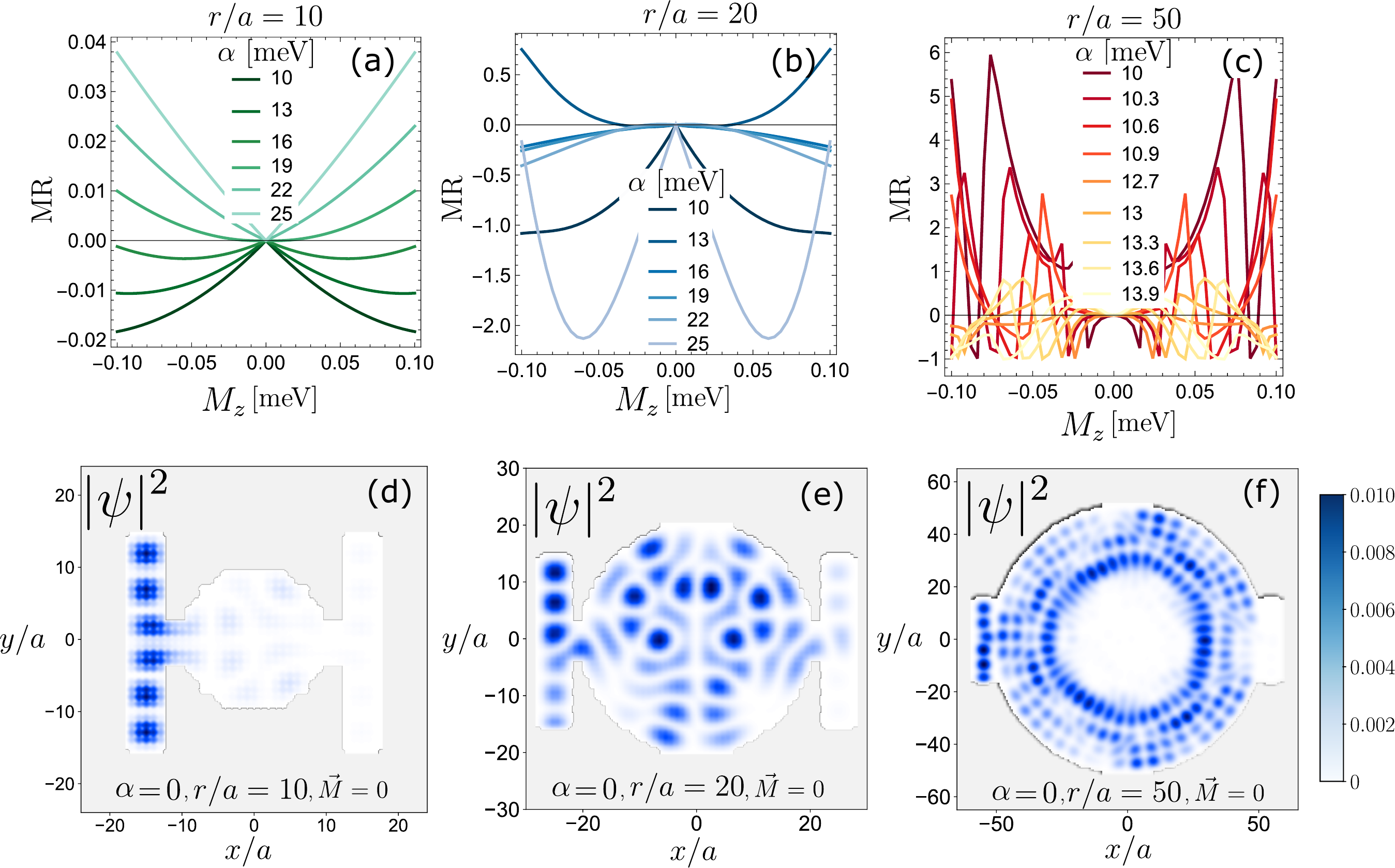}
	\caption{(a)–(c) Magnetoresistance MR as a function of $M_z$, for increasing values of the dot radius $r = 10$, 20, and 50, and several Rashba coupling strengths $\alpha$. For small $r$, MR shows a clear crossover from WL to WAL as $\alpha$ increases. As $r$ grows, this transition is suppressed, and the MR response becomes more complex. (d)–(f) Probability densities $|\psi|^2$ of the scattering wavefunction for $\alpha = 0$ and $\vec{M} = 0$, with an electron injected from lead 1. These panels represent single scattering processes and illustrate how the spatial distribution of the wavefunction changes with $r$. For $r = 10$, backscattering dominates, consistent with the WL regime. As $r$ increases, transmission becomes more prominent.}
	\label{FigureApp1}
\end{figure*}
\section{Effect of quantum dot size on interference-induced magnetoresistance}
\label{AppA}
In Fig.~\ref{FigureApp1}(a)-(c), we show MR as a function of out-of-plane magnetization \( M_z \) for increasing quantum dot radius \( r = 10 \), \( 20 \), and \( 50 \). For \( r = 10 \), we observe a clear transition from WL to WAL as \( \alpha \) increases (Fig.~\ref{FigureApp1}(a)). As the dot radius increases, the MR curves become more irregular and the characteristic WL–WAL crossover gradually fades [Fig.~\ref{FigureApp1}(b)-(c)]. The behavior at \( r = 10 \) is associated with a kinematic regime where the Fermi wavelength \( \lambda_F = 2\pi/k_F \) is comparable to the dot radius. In this limit, multiple coherent scattering processes form closed clockwise and counterclockwise paths that interfere constructively, enhancing the reflection probability. This interference pattern is visible in the scattering wavefunction density \( |\psi|^2 \) at \( \alpha = 0 \) shown in Fig.~\ref{FigureApp1}(d). For larger radii [Figs.~\ref{FigureApp1}(e)-(f)], the condition \( r \sim \lambda_F \) no longer holds. As a result, the constructive interference responsible for WL is suppressed, and the wavefunction profiles indicate increased transmission through the structure, Figs.~\ref{FigureApp1}(e)-(f).
\section{Suppression of spurious spin currents}
\label{AppB}
\begin{figure}[!htbp]
	\centering
	\includegraphics[width=0.4\textwidth]{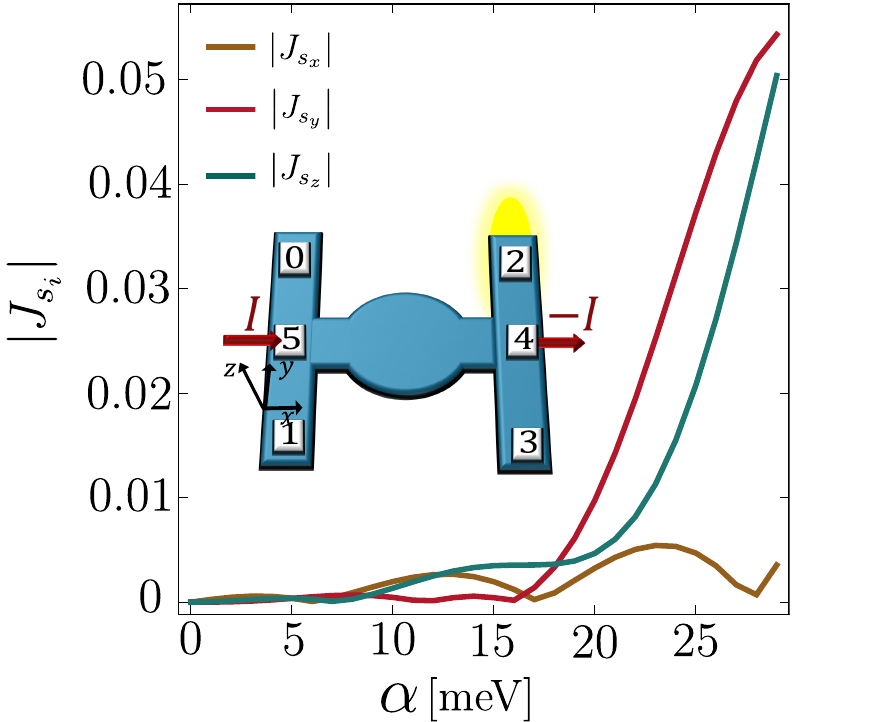}
	\caption{ Absolute values of the $x$-, $y$-, and $z$-polarized spin currents in lead 2 as a function of $\alpha$, for a six-terminal geometry with current injection along the $x$-direction. In this setup, the spurious Edelstein-like contribution to $J_{s_x}$ is suppressed, in contrast to the four-terminal configuration in Fig. 2. The $J_{s_i}$ components are expressed in units of $J^0_s=e V_0/4 \pi$, with $V_0=h I/e^2$ and $I$ is the applied current bias.}
	\label{FigureApp2}
\end{figure}

In Fig.~\ref{FigureApp2}, we show that, by adopting a suitable bias configuration involving six terminals and injecting a charge current along the \( x \)-direction, it is possible to selectively suppress unwanted spin current components. In particular, the spurious in-plane spin current \( |J_{s_x}| \), which can arise due to confinement effects, is fully suppressed across all values of the Rashba coupling \( \alpha\). As a result, only the physically meaningful spin responses remain: the Edelstein current \( |J_{s_y}| \), which is associated with spin polarization along \( y \) induced by the applied bias, and the transverse spin Hall current \( |J_{s_z}| \), which flows perpendicular to the injected current. Both \( |J_{s_y}| \) and \( |J_{s_z}| \) grow with increasing \( \alpha \), reflecting the intrinsic spin-charge conversion mechanisms activated by the Rashba interaction. This setup thus enables a clean separation of spin current contributions arising from charge-to-spin conversion mechanisms.
\section{Rashba Hamiltonian with in-plane Zeeman field}
\label{AppC}
\begin{figure*}
	\centering
	\includegraphics[width=0.8\textwidth]{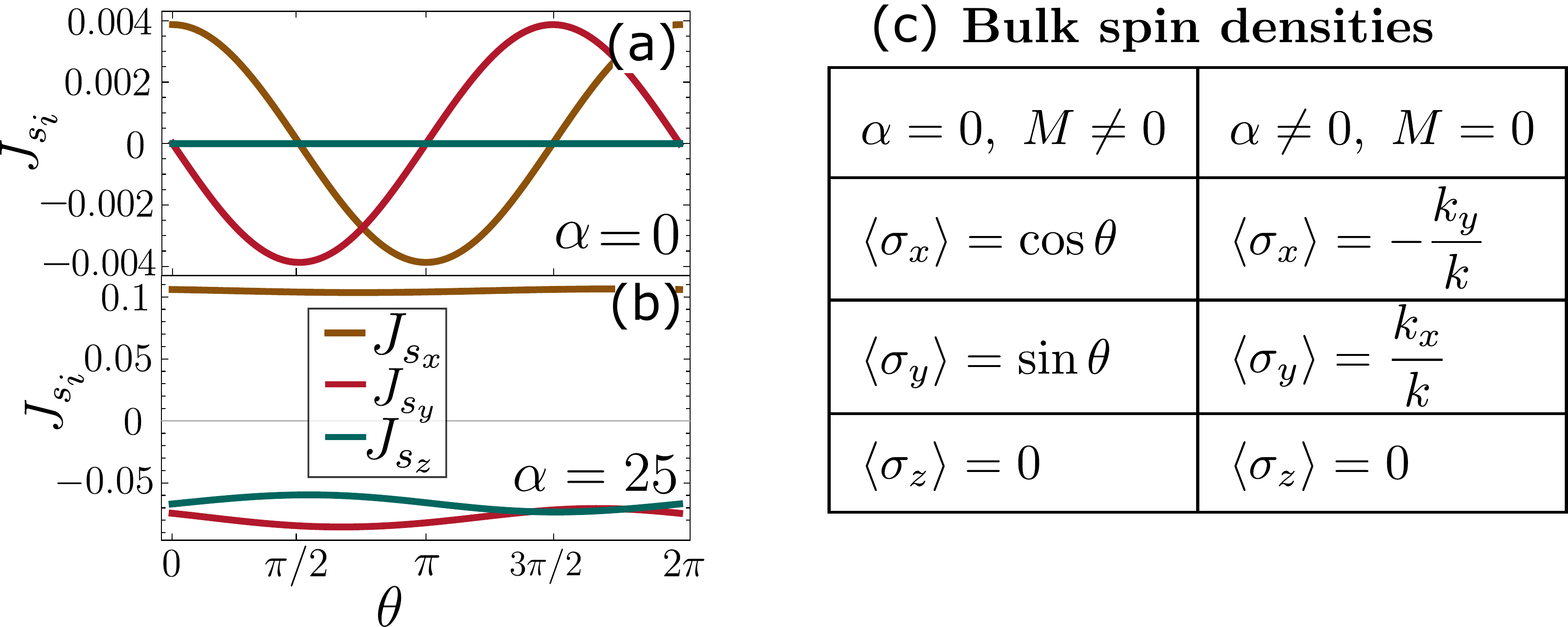}
	\caption{$x$, $y$ and $z$-components of the spin current obtained for $\alpha=0$, panel (a), and $\alpha=25$, panel (b). (c) Expectation values of spin density operators over the lowest band eigenstate of a bulk Hamiltonian with the Rashba coupling $\alpha$ and Zeeman energy $M$, induced by an in-plane magnetic field. The $J_{s_i}$ components are expressed in units of $J^0_s=e V_0/4 \pi$, with $V_0=2 \pi \hbar I/e^2$ and $I$ is the applied current bias.}
	\label{FigureApp3}
\end{figure*}
We consider the single-particle Hamiltonian in momentum space corresponding to the continuum limit of the tight-binding model discussed in the main text. The Hamiltonian for a two-dimensional electron gas with Rashba spin–orbit coupling and an in-plane Zeeman field is given by
\begin{equation}
	H(\vec{k}) = \frac{\hbar^2 k^2}{2m}\sigma_0 + \alpha_R(\sigma_x k_y - \sigma_y k_x) + M(\cos\theta \sigma_x + \sin\theta \sigma_y).
	\label{RashbaHk}
\end{equation}
Here $\alpha_R=2 a \alpha$ is the continuum Rashba coefficient associated with the tight-binding Rashba parameter $\alpha$ and the lattice constat $a$ introduced in the main text, while $M$ and $\theta$ denote respectively the modulus and the direction of the in-plane Zeeman field. We explicitly notice that for $\alpha_R=0$ the unitary equivalence $\sigma_y\ H(\theta=0,\ \alpha_R=0)\ \sigma_y=H(\theta=\pi,\ \alpha_R=0)$ can be verified, where $H(\theta,\alpha_R)$ refers to Eq.~(\ref{RashbaHk}).\\
For $\alpha_R \neq 0$ one observes that $\sigma_y\ H(\theta=0,\ \alpha_R \neq 0)\ \sigma_y \neq H(\theta=\pi,\ \alpha_R\neq 0)$, being this observation related to the interpretation of Fig.~\ref{Figure3} of the main text. The Hamiltonian can be written in compact form as:
\begin{equation}
	H(\vec{k}) = \frac{\hbar^2 k^2}{2m} \sigma_0 + \vec{h}(\vec{k}) \cdot \vec{\sigma},
\end{equation}
with effective field:
\begin{equation}
	\vec{h}(\vec{k}) = 
	\begin{pmatrix}
		\alpha_R k_y + M \cos\theta \\
		- \alpha_R k_x + M \sin\theta \\
		0
	\end{pmatrix}.
\end{equation}
The eigenvalues of $H(\vec{k})$ are:
\begin{equation}
	E_\pm(\vec{k}) = \frac{\hbar^2 k^2}{2m} \pm |\vec{h}(\vec{k})|,
\end{equation}
where:
\begin{equation}
	|\vec{h}(\vec{k})| = \sqrt{(\alpha_R k_y + M \cos\theta)^2 + (-\alpha_R k_x + M \sin\theta)^2}.
\end{equation}
The corresponding normalized eigenstates are:
\begin{equation}
	|\vec{k}, \pm \rangle = \frac{1}{\sqrt{2}} \begin{pmatrix}
		\pm 1 \\
		e^{-i \phi_h}
	\end{pmatrix}, \quad \text{with} \quad \phi_h = \arg(h_x + i h_y),
\end{equation}
where $h_x = \alpha_R k_y + M \cos\theta$ and $h_y = -\alpha_R k_x + M \sin\theta$. The expectation values of the Pauli matrices on these states are:
\begin{equation}
	\langle \vec{k}, \pm | \vec{\sigma} | \vec{k}, \pm \rangle =
	\pm \frac{1}{|\vec{h}(\vec{k})|}
	\begin{pmatrix}
		\alpha_R k_y + M \cos\theta \\
		- \alpha_R k_x + M \sin\theta \\
		0
	\end{pmatrix}.
\end{equation}
These results describe spin-momentum locking in the presence of both spin-orbit and Zeeman interactions, with spin orientation aligned or anti-aligned to the effective in-plane field $\vec{h}(\vec{k})$.\\
In the absence of spin-orbit coupling ($\alpha_R = 0$), the spin aligns with the in-plane Zeeman field, yielding $\langle \vec{\sigma} \rangle = (\cos\theta, \sin\theta, 0)$ for the lower-energy state. Conversely, in the absence of Zeeman field ($M = 0$), the spin lies orthogonal to the momentum due to Rashba interaction: $\langle \vec{\sigma} \rangle = (-k_y/k, k_x/k, 0)$. These two limits are summarized in Fig.~\ref{FigureApp3}(c) and are recovered from the general expressions derived above. This behaviour is consistent with the spin current behaviour shown in Figs. \ref{FigureApp3}(a)-(b), since spin current and spin density are roughly proportional in the low-filling regime.
\bibliographystyle{apsrev}
\bibliography{Bib}

\begin{thebibliography}{46}
\expandafter\ifx\csname natexlab\endcsname\relax\def\natexlab#1{#1}\fi
\expandafter\ifx\csname bibnamefont\endcsname\relax
  \def\bibnamefont#1{#1}\fi
\expandafter\ifx\csname bibfnamefont\endcsname\relax
  \def\bibfnamefont#1{#1}\fi
\expandafter\ifx\csname citenamefont\endcsname\relax
  \def\citenamefont#1{#1}\fi
\expandafter\ifx\csname url\endcsname\relax
  \def\url#1{\texttt{#1}}\fi
\expandafter\ifx\csname urlprefix\endcsname\relax\def\urlprefix{URL }\fi
\providecommand{\bibinfo}[2]{#2}
\providecommand{\eprint}[2][]{\url{#2}}

\bibitem[{\citenamefont{Kurdak et~al.}(1992)\citenamefont{Kurdak, Chang, Chin,
  and Chang}}]{PhysRevB.46.6846}
\bibinfo{author}{\bibfnamefont{I.~M.~C.} \bibnamefont{Kurdak}},
  \bibinfo{author}{\bibfnamefont{A.~M.} \bibnamefont{Chang}},
  \bibinfo{author}{\bibfnamefont{A.}~\bibnamefont{Chin}}, \bibnamefont{and}
  \bibinfo{author}{\bibfnamefont{T.~Y.} \bibnamefont{Chang}},
  \bibinfo{journal}{Phys. Rev. B} \textbf{\bibinfo{volume}{46}},
  \bibinfo{pages}{6846} (\bibinfo{year}{1992}),
  \urlprefix\url{https://link.aps.org/doi/10.1103/PhysRevB.46.6846}.

\bibitem[{\citenamefont{Hirayama et~al.}(2011)\citenamefont{Hirayama, Aoki, and
  Kato}}]{PhysRevLett.107.027204}
\bibinfo{author}{\bibfnamefont{H.}~\bibnamefont{Hirayama}},
  \bibinfo{author}{\bibfnamefont{Y.}~\bibnamefont{Aoki}}, \bibnamefont{and}
  \bibinfo{author}{\bibfnamefont{C.}~\bibnamefont{Kato}},
  \bibinfo{journal}{Phys. Rev. Lett.} \textbf{\bibinfo{volume}{107}},
  \bibinfo{pages}{027204} (\bibinfo{year}{2011}),
  \urlprefix\url{https://link.aps.org/doi/10.1103/PhysRevLett.107.027204}.

\bibitem[{\citenamefont{Chen et~al.}(2025)\citenamefont{Chen, D'Antuono, Trama,
  Preziosi, Jouault, Teppe, Consejo, Perroni, Citro, Stornaiuolo
  et~al.}}]{Trama}
\bibinfo{author}{\bibfnamefont{Y.}~\bibnamefont{Chen}},
  \bibinfo{author}{\bibfnamefont{M.}~\bibnamefont{D'Antuono}},
  \bibinfo{author}{\bibfnamefont{M.}~\bibnamefont{Trama}},
  \bibinfo{author}{\bibfnamefont{D.}~\bibnamefont{Preziosi}},
  \bibinfo{author}{\bibfnamefont{B.}~\bibnamefont{Jouault}},
  \bibinfo{author}{\bibfnamefont{F.}~\bibnamefont{Teppe}},
  \bibinfo{author}{\bibfnamefont{C.}~\bibnamefont{Consejo}},
  \bibinfo{author}{\bibfnamefont{C.~A.} \bibnamefont{Perroni}},
  \bibinfo{author}{\bibfnamefont{R.}~\bibnamefont{Citro}},
  \bibinfo{author}{\bibfnamefont{D.}~\bibnamefont{Stornaiuolo}},
  \bibnamefont{et~al.}, \bibinfo{journal}{Advanced Materials}
  \textbf{\bibinfo{volume}{37}}, \bibinfo{pages}{2410354}
  (\bibinfo{year}{2025}),
  \urlprefix\url{https://advanced.onlinelibrary.wiley.com/doi/abs/10.1002/adma.202410354}.

\bibitem[{\citenamefont{Liang et~al.}(2009)\citenamefont{Liang, Yang, Wang, and
  Chan}}]{Liang_2009}
\bibinfo{author}{\bibfnamefont{F.}~\bibnamefont{Liang}},
  \bibinfo{author}{\bibfnamefont{Y.~H.} \bibnamefont{Yang}},
  \bibinfo{author}{\bibfnamefont{J.}~\bibnamefont{Wang}}, \bibnamefont{and}
  \bibinfo{author}{\bibfnamefont{K.~S.} \bibnamefont{Chan}},
  \bibinfo{journal}{Europhysics Letters} \textbf{\bibinfo{volume}{87}},
  \bibinfo{pages}{47004} (\bibinfo{year}{2009}),
  \urlprefix\url{https://dx.doi.org/10.1209/0295-5075/87/47004}.

\bibitem[{\citenamefont{Nitta and Koga}(2003)}]{Nitta2003}
\bibinfo{author}{\bibfnamefont{J.}~\bibnamefont{Nitta}} \bibnamefont{and}
  \bibinfo{author}{\bibfnamefont{T.}~\bibnamefont{Koga}},
  \bibinfo{journal}{Journal of Superconductivity}
  \textbf{\bibinfo{volume}{16}}, \bibinfo{pages}{689} (\bibinfo{year}{2003}),
  \urlprefix\url{https://doi.org/10.1023/A:1025309805995}.

\bibitem[{\citenamefont{{Bychkov} and {Rashba}}(1984)}]{Bychkov1984}
\bibinfo{author}{\bibfnamefont{Y.~A.} \bibnamefont{{Bychkov}}}
  \bibnamefont{and} \bibinfo{author}{\bibfnamefont{{\'E}.~I.}
  \bibnamefont{{Rashba}}}, \bibinfo{journal}{ZhETF Pisma Redaktsiiu}
  \textbf{\bibinfo{volume}{39}}, \bibinfo{pages}{66} (\bibinfo{year}{1984}),
  \urlprefix\url{https://ui.adsabs.harvard.edu/abs/1984ZhPmR..39...66B}.

\bibitem[{\citenamefont{Awschalom and Flatté}(2007)}]{Awschalom2007}
\bibinfo{author}{\bibfnamefont{D.}~\bibnamefont{Awschalom}} \bibnamefont{and}
  \bibinfo{author}{\bibfnamefont{M.}~\bibnamefont{Flatté}},
  \bibinfo{journal}{Nature Physics} \textbf{\bibinfo{volume}{3}},
  \bibinfo{pages}{153} (\bibinfo{year}{2007}),
  \urlprefix\url{https://doi.org/10.1038/nphys551}.

\bibitem[{\citenamefont{Lutchyn et~al.}(2010)\citenamefont{Lutchyn, Sau, and
  Das~Sarma}}]{Lutchyn2010}
\bibinfo{author}{\bibfnamefont{R.~M.} \bibnamefont{Lutchyn}},
  \bibinfo{author}{\bibfnamefont{J.~D.} \bibnamefont{Sau}}, \bibnamefont{and}
  \bibinfo{author}{\bibfnamefont{S.}~\bibnamefont{Das~Sarma}},
  \bibinfo{journal}{Phys. Rev. Lett.} \textbf{\bibinfo{volume}{105}},
  \bibinfo{pages}{077001} (\bibinfo{year}{2010}),
  \urlprefix\url{https://doi.org/10.1103/PhysRevLett.105.077001}.

\bibitem[{\citenamefont{Alicea}(2012)}]{Alicea2012}
\bibinfo{author}{\bibfnamefont{J.}~\bibnamefont{Alicea}},
  \bibinfo{journal}{Reports on Progress in Physics}
  \textbf{\bibinfo{volume}{75}}, \bibinfo{pages}{076501}
  (\bibinfo{year}{2012}),
  \urlprefix\url{https://doi.org/10.1088/0034-4885/75/7/076501}.

\bibitem[{\citenamefont{Yokoyama et~al.}(2009)\citenamefont{Yokoyama, Tanaka,
  and Nagaosa}}]{Yokoyama2009}
\bibinfo{author}{\bibfnamefont{T.}~\bibnamefont{Yokoyama}},
  \bibinfo{author}{\bibfnamefont{Y.}~\bibnamefont{Tanaka}}, \bibnamefont{and}
  \bibinfo{author}{\bibfnamefont{N.}~\bibnamefont{Nagaosa}},
  \bibinfo{journal}{Phys. Rev. B} \textbf{\bibinfo{volume}{80}},
  \bibinfo{pages}{125339} (\bibinfo{year}{2009}),
  \urlprefix\url{https://doi.org/10.1103/PhysRevB.80.125339}.

\bibitem[{\citenamefont{Maiellaro et~al.}(2024)\citenamefont{Maiellaro, Trama,
  Settino, Guarcello, Romeo, and Citro}}]{SciPostPhysMa}
\bibinfo{author}{\bibfnamefont{A.}~\bibnamefont{Maiellaro}},
  \bibinfo{author}{\bibfnamefont{M.}~\bibnamefont{Trama}},
  \bibinfo{author}{\bibfnamefont{J.}~\bibnamefont{Settino}},
  \bibinfo{author}{\bibfnamefont{C.}~\bibnamefont{Guarcello}},
  \bibinfo{author}{\bibfnamefont{F.}~\bibnamefont{Romeo}}, \bibnamefont{and}
  \bibinfo{author}{\bibfnamefont{R.}~\bibnamefont{Citro}},
  \bibinfo{journal}{SciPost Phys.} \textbf{\bibinfo{volume}{17}},
  \bibinfo{pages}{101} (\bibinfo{year}{2024}),
  \urlprefix\url{https://scipost.org/10.21468/SciPostPhys.17.4.101}.

\bibitem[{\citenamefont{Guarcello et~al.}(2024)\citenamefont{Guarcello,
  Maiellaro, Settino, Gaiardoni, Trama, Romeo, and Citro}}]{GUARCELLO}
\bibinfo{author}{\bibfnamefont{C.}~\bibnamefont{Guarcello}},
  \bibinfo{author}{\bibfnamefont{A.}~\bibnamefont{Maiellaro}},
  \bibinfo{author}{\bibfnamefont{J.}~\bibnamefont{Settino}},
  \bibinfo{author}{\bibfnamefont{I.}~\bibnamefont{Gaiardoni}},
  \bibinfo{author}{\bibfnamefont{M.}~\bibnamefont{Trama}},
  \bibinfo{author}{\bibfnamefont{F.}~\bibnamefont{Romeo}}, \bibnamefont{and}
  \bibinfo{author}{\bibfnamefont{R.}~\bibnamefont{Citro}},
  \bibinfo{journal}{Chaos, Solitons \& Fractals}
  \textbf{\bibinfo{volume}{189}}, \bibinfo{pages}{115596}
  (\bibinfo{year}{2024}), ISSN \bibinfo{issn}{0960-0779},
  \urlprefix\url{https://www.sciencedirect.com/science/article/pii/S0960077924011482}.

\bibitem[{\citenamefont{Li et~al.}(2018)\citenamefont{Li, Liu, Wu, Li, Ren, and
  Li}}]{Li2018}
\bibinfo{author}{\bibfnamefont{R.}~\bibnamefont{Li}},
  \bibinfo{author}{\bibfnamefont{Z.-H.} \bibnamefont{Liu}},
  \bibinfo{author}{\bibfnamefont{Y.}~\bibnamefont{Wu}},
  \bibinfo{author}{\bibfnamefont{Z.-Q.} \bibnamefont{Li}},
  \bibinfo{author}{\bibfnamefont{J.}~\bibnamefont{Ren}}, \bibnamefont{and}
  \bibinfo{author}{\bibfnamefont{Y.-Q.} \bibnamefont{Li}},
  \bibinfo{journal}{Scientific Reports} \textbf{\bibinfo{volume}{8}},
  \bibinfo{pages}{7400} (\bibinfo{year}{2018}),
  \urlprefix\url{https://doi.org/10.1038/s41598-018-25692-2}.

\bibitem[{\citenamefont{Aleiner and Fal'ko}(2001)}]{PhysRevLett.87.256801}
\bibinfo{author}{\bibfnamefont{I.~L.} \bibnamefont{Aleiner}} \bibnamefont{and}
  \bibinfo{author}{\bibfnamefont{V.~I.} \bibnamefont{Fal'ko}},
  \bibinfo{journal}{Phys. Rev. Lett.} \textbf{\bibinfo{volume}{87}},
  \bibinfo{pages}{256801} (\bibinfo{year}{2001}),
  \urlprefix\url{https://link.aps.org/doi/10.1103/PhysRevLett.87.256801}.

\bibitem[{\citenamefont{Maiellaro et~al.}(2025)\citenamefont{Maiellaro, Romeo,
  Trama, Gaiardoni, Settino, Guarcello, Bergeal, Bibes, and
  Citro}}]{Maiellaro2025}
\bibinfo{author}{\bibfnamefont{A.}~\bibnamefont{Maiellaro}},
  \bibinfo{author}{\bibfnamefont{F.}~\bibnamefont{Romeo}},
  \bibinfo{author}{\bibfnamefont{M.}~\bibnamefont{Trama}},
  \bibinfo{author}{\bibfnamefont{I.}~\bibnamefont{Gaiardoni}},
  \bibinfo{author}{\bibfnamefont{J.}~\bibnamefont{Settino}},
  \bibinfo{author}{\bibfnamefont{C.}~\bibnamefont{Guarcello}},
  \bibinfo{author}{\bibfnamefont{N.}~\bibnamefont{Bergeal}},
  \bibinfo{author}{\bibfnamefont{M.}~\bibnamefont{Bibes}}, \bibnamefont{and}
  \bibinfo{author}{\bibfnamefont{R.}~\bibnamefont{Citro}},
  \bibinfo{journal}{Phys. Rev. Res.} \textbf{\bibinfo{volume}{7}},
  \bibinfo{pages}{043100} (\bibinfo{year}{2025}),
  \urlprefix\url{https://link.aps.org/doi/10.1103/zpvy-t4d4}.

\bibitem[{\citenamefont{Zucchetti et~al.}(2018)\citenamefont{Zucchetti, Dau,
  Bottegoni, Vergnaud, Guillet, Marty, Beign\'e, Gambarelli, Picone, Calloni
  et~al.}}]{PhysRevB.98.184418}
\bibinfo{author}{\bibfnamefont{C.}~\bibnamefont{Zucchetti}},
  \bibinfo{author}{\bibfnamefont{M.-T.} \bibnamefont{Dau}},
  \bibinfo{author}{\bibfnamefont{F.}~\bibnamefont{Bottegoni}},
  \bibinfo{author}{\bibfnamefont{C.}~\bibnamefont{Vergnaud}},
  \bibinfo{author}{\bibfnamefont{T.}~\bibnamefont{Guillet}},
  \bibinfo{author}{\bibfnamefont{A.}~\bibnamefont{Marty}},
  \bibinfo{author}{\bibfnamefont{C.}~\bibnamefont{Beign\'e}},
  \bibinfo{author}{\bibfnamefont{S.}~\bibnamefont{Gambarelli}},
  \bibinfo{author}{\bibfnamefont{A.}~\bibnamefont{Picone}},
  \bibinfo{author}{\bibfnamefont{A.}~\bibnamefont{Calloni}},
  \bibnamefont{et~al.}, \bibinfo{journal}{Phys. Rev. B}
  \textbf{\bibinfo{volume}{98}}, \bibinfo{pages}{184418}
  (\bibinfo{year}{2018}),
  \urlprefix\url{https://link.aps.org/doi/10.1103/PhysRevB.98.184418}.

\bibitem[{\citenamefont{Hanlon et~al.}(2021)\citenamefont{Hanlon, Oberg, Chen,
  and Doherty}}]{PhysRevApplied}
\bibinfo{author}{\bibfnamefont{L.}~\bibnamefont{Hanlon}},
  \bibinfo{author}{\bibfnamefont{L.}~\bibnamefont{Oberg}},
  \bibinfo{author}{\bibfnamefont{Y.}~\bibnamefont{Chen}}, \bibnamefont{and}
  \bibinfo{author}{\bibfnamefont{M.~W.} \bibnamefont{Doherty}},
  \bibinfo{journal}{Phys. Rev. Appl.} \textbf{\bibinfo{volume}{16}},
  \bibinfo{pages}{064050} (\bibinfo{year}{2021}),
  \urlprefix\url{https://link.aps.org/doi/10.1103/PhysRevApplied.16.064050}.

\bibitem[{\citenamefont{Miller et~al.}(2003)\citenamefont{Miller, Zumb\"uhl,
  Marcus, Lyanda-Geller, Goldhaber-Gordon, Campman, and
  Gossard}}]{PhysRevLett.90.076807}
\bibinfo{author}{\bibfnamefont{J.~B.} \bibnamefont{Miller}},
  \bibinfo{author}{\bibfnamefont{D.~M.} \bibnamefont{Zumb\"uhl}},
  \bibinfo{author}{\bibfnamefont{C.~M.} \bibnamefont{Marcus}},
  \bibinfo{author}{\bibfnamefont{Y.~B.} \bibnamefont{Lyanda-Geller}},
  \bibinfo{author}{\bibfnamefont{D.}~\bibnamefont{Goldhaber-Gordon}},
  \bibinfo{author}{\bibfnamefont{K.}~\bibnamefont{Campman}}, \bibnamefont{and}
  \bibinfo{author}{\bibfnamefont{A.~C.} \bibnamefont{Gossard}},
  \bibinfo{journal}{Phys. Rev. Lett.} \textbf{\bibinfo{volume}{90}},
  \bibinfo{pages}{076807} (\bibinfo{year}{2003}),
  \urlprefix\url{https://link.aps.org/doi/10.1103/PhysRevLett.90.076807}.

\bibitem[{\citenamefont{Zumb\"uhl et~al.}(2002)\citenamefont{Zumb\"uhl, Miller,
  Marcus, Campman, and Gossard}}]{PhysRevLett.89.276803}
\bibinfo{author}{\bibfnamefont{D.~M.} \bibnamefont{Zumb\"uhl}},
  \bibinfo{author}{\bibfnamefont{J.~B.} \bibnamefont{Miller}},
  \bibinfo{author}{\bibfnamefont{C.~M.} \bibnamefont{Marcus}},
  \bibinfo{author}{\bibfnamefont{K.}~\bibnamefont{Campman}}, \bibnamefont{and}
  \bibinfo{author}{\bibfnamefont{A.~C.} \bibnamefont{Gossard}},
  \bibinfo{journal}{Phys. Rev. Lett.} \textbf{\bibinfo{volume}{89}},
  \bibinfo{pages}{276803} (\bibinfo{year}{2002}),
  \urlprefix\url{https://link.aps.org/doi/10.1103/PhysRevLett.89.276803}.

\bibitem[{\citenamefont{Trier et~al.}(2020)\citenamefont{Trier, Vaz, Bruneel,
  Noël, Fert, Vila, Attan{\'e}, Barth{\'e}l{\'e}my, Gabay, Jaffrès
  et~al.}}]{nanolett.9b04079}
\bibinfo{author}{\bibfnamefont{F.}~\bibnamefont{Trier}},
  \bibinfo{author}{\bibfnamefont{D.~C.} \bibnamefont{Vaz}},
  \bibinfo{author}{\bibfnamefont{P.}~\bibnamefont{Bruneel}},
  \bibinfo{author}{\bibfnamefont{P.}~\bibnamefont{Noël}},
  \bibinfo{author}{\bibfnamefont{A.}~\bibnamefont{Fert}},
  \bibinfo{author}{\bibfnamefont{L.}~\bibnamefont{Vila}},
  \bibinfo{author}{\bibfnamefont{J.-P.} \bibnamefont{Attan{\'e}}},
  \bibinfo{author}{\bibfnamefont{A.}~\bibnamefont{Barth{\'e}l{\'e}my}},
  \bibinfo{author}{\bibfnamefont{M.}~\bibnamefont{Gabay}},
  \bibinfo{author}{\bibfnamefont{H.}~\bibnamefont{Jaffrès}},
  \bibnamefont{et~al.}, \bibinfo{journal}{Nano Letters}
  \textbf{\bibinfo{volume}{20}}, \bibinfo{pages}{395} (\bibinfo{year}{2020}),
  \bibinfo{note}{pMID: 31859513},
  \eprint{https://doi.org/10.1021/acs.nanolett.9b04079},
  \urlprefix\url{https://doi.org/10.1021/acs.nanolett.9b04079}.

\bibitem[{\citenamefont{Edelstein}(1990)}]{EDELSTEIN1990233}
\bibinfo{author}{\bibfnamefont{V.}~\bibnamefont{Edelstein}},
  \bibinfo{journal}{Solid State Communications} \textbf{\bibinfo{volume}{73}},
  \bibinfo{pages}{233} (\bibinfo{year}{1990}), ISSN \bibinfo{issn}{0038-1098},
  \urlprefix\url{https://www.sciencedirect.com/science/article/pii/003810989090963C}.

\bibitem[{\citenamefont{Sinova et~al.}(2004)\citenamefont{Sinova, Culcer, Niu,
  Sinitsyn, Jungwirth, and MacDonald}}]{Sinova2004}
\bibinfo{author}{\bibfnamefont{J.}~\bibnamefont{Sinova}},
  \bibinfo{author}{\bibfnamefont{D.}~\bibnamefont{Culcer}},
  \bibinfo{author}{\bibfnamefont{Q.}~\bibnamefont{Niu}},
  \bibinfo{author}{\bibfnamefont{N.~A.} \bibnamefont{Sinitsyn}},
  \bibinfo{author}{\bibfnamefont{T.}~\bibnamefont{Jungwirth}},
  \bibnamefont{and} \bibinfo{author}{\bibfnamefont{A.~H.}
  \bibnamefont{MacDonald}}, \bibinfo{journal}{Phys. Rev. Lett.}
  \textbf{\bibinfo{volume}{92}}, \bibinfo{pages}{126603}
  (\bibinfo{year}{2004}),
  \urlprefix\url{https://link.aps.org/doi/10.1103/PhysRevLett.92.126603}.

\bibitem[{\citenamefont{Aharonov and Bohm}(1959)}]{PhysRev.115.485}
\bibinfo{author}{\bibfnamefont{Y.}~\bibnamefont{Aharonov}} \bibnamefont{and}
  \bibinfo{author}{\bibfnamefont{D.}~\bibnamefont{Bohm}},
  \bibinfo{journal}{Phys. Rev.} \textbf{\bibinfo{volume}{115}},
  \bibinfo{pages}{485} (\bibinfo{year}{1959}),
  \urlprefix\url{https://link.aps.org/doi/10.1103/PhysRev.115.485}.

\bibitem[{\citenamefont{Bergmann}(1984{\natexlab{a}})}]{Bergmann1984}
\bibinfo{author}{\bibfnamefont{G.}~\bibnamefont{Bergmann}},
  \bibinfo{journal}{Physics Reports} \textbf{\bibinfo{volume}{107}},
  \bibinfo{pages}{1} (\bibinfo{year}{1984}{\natexlab{a}}), ISSN
  \bibinfo{issn}{0370-1573},
  \urlprefix\url{https://www.sciencedirect.com/science/article/pii/0370157384901030}.

\bibitem[{\citenamefont{Hikami et~al.}(1980)\citenamefont{Hikami, Larkin, and
  Nagaoka}}]{Hikami1980}
\bibinfo{author}{\bibfnamefont{S.}~\bibnamefont{Hikami}},
  \bibinfo{author}{\bibfnamefont{A.~I.} \bibnamefont{Larkin}},
  \bibnamefont{and} \bibinfo{author}{\bibfnamefont{Y.}~\bibnamefont{Nagaoka}},
  \bibinfo{journal}{Progress of Theoretical Physics}
  \textbf{\bibinfo{volume}{63}}, \bibinfo{pages}{707} (\bibinfo{year}{1980}),
  ISSN \bibinfo{issn}{0033-068X},
  \eprint{https://academic.oup.com/ptp/article-pdf/63/2/707/5336056/63-2-707.pdf},
  \urlprefix\url{https://doi.org/10.1143/PTP.63.707}.

\bibitem[{\citenamefont{Bergmann}(1984{\natexlab{b}})}]{PhysRevLett.53.1100}
\bibinfo{author}{\bibfnamefont{G.}~\bibnamefont{Bergmann}},
  \bibinfo{journal}{Phys. Rev. Lett.} \textbf{\bibinfo{volume}{53}},
  \bibinfo{pages}{1100} (\bibinfo{year}{1984}{\natexlab{b}}),
  \urlprefix\url{https://link.aps.org/doi/10.1103/PhysRevLett.53.1100}.

\bibitem[{\citenamefont{Koga et~al.}(2002)\citenamefont{Koga, Nitta, Akazaki,
  and Takayanagi}}]{PhysRevLett.89.046801}
\bibinfo{author}{\bibfnamefont{T.}~\bibnamefont{Koga}},
  \bibinfo{author}{\bibfnamefont{J.}~\bibnamefont{Nitta}},
  \bibinfo{author}{\bibfnamefont{T.}~\bibnamefont{Akazaki}}, \bibnamefont{and}
  \bibinfo{author}{\bibfnamefont{H.}~\bibnamefont{Takayanagi}},
  \bibinfo{journal}{Phys. Rev. Lett.} \textbf{\bibinfo{volume}{89}},
  \bibinfo{pages}{046801} (\bibinfo{year}{2002}),
  \urlprefix\url{https://link.aps.org/doi/10.1103/PhysRevLett.89.046801}.

\bibitem[{\citenamefont{Steinberg et~al.}(2011)\citenamefont{Steinberg,
  Lalo\"e, Fatemi, Moodera, and Jarillo-Herrero}}]{PhysRevB.84.233101}
\bibinfo{author}{\bibfnamefont{H.}~\bibnamefont{Steinberg}},
  \bibinfo{author}{\bibfnamefont{J.-B.} \bibnamefont{Lalo\"e}},
  \bibinfo{author}{\bibfnamefont{V.}~\bibnamefont{Fatemi}},
  \bibinfo{author}{\bibfnamefont{J.~S.} \bibnamefont{Moodera}},
  \bibnamefont{and}
  \bibinfo{author}{\bibfnamefont{P.}~\bibnamefont{Jarillo-Herrero}},
  \bibinfo{journal}{Phys. Rev. B} \textbf{\bibinfo{volume}{84}},
  \bibinfo{pages}{233101} (\bibinfo{year}{2011}),
  \urlprefix\url{https://link.aps.org/doi/10.1103/PhysRevB.84.233101}.

\bibitem[{\citenamefont{Chen et~al.}(2010)\citenamefont{Chen, Qin, Yang, Liu,
  Guan, Qu, Zhang, Shi, Xie, Yang et~al.}}]{PhysRevLett.105.176602}
\bibinfo{author}{\bibfnamefont{J.}~\bibnamefont{Chen}},
  \bibinfo{author}{\bibfnamefont{H.~J.} \bibnamefont{Qin}},
  \bibinfo{author}{\bibfnamefont{F.}~\bibnamefont{Yang}},
  \bibinfo{author}{\bibfnamefont{J.}~\bibnamefont{Liu}},
  \bibinfo{author}{\bibfnamefont{T.}~\bibnamefont{Guan}},
  \bibinfo{author}{\bibfnamefont{F.~M.} \bibnamefont{Qu}},
  \bibinfo{author}{\bibfnamefont{G.~H.} \bibnamefont{Zhang}},
  \bibinfo{author}{\bibfnamefont{J.~R.} \bibnamefont{Shi}},
  \bibinfo{author}{\bibfnamefont{X.~C.} \bibnamefont{Xie}},
  \bibinfo{author}{\bibfnamefont{C.~L.} \bibnamefont{Yang}},
  \bibnamefont{et~al.}, \bibinfo{journal}{Phys. Rev. Lett.}
  \textbf{\bibinfo{volume}{105}}, \bibinfo{pages}{176602}
  (\bibinfo{year}{2010}),
  \urlprefix\url{https://link.aps.org/doi/10.1103/PhysRevLett.105.176602}.

\bibitem[{\citenamefont{Caviglia et~al.}(2010)\citenamefont{Caviglia, Gabay,
  Gariglio, Reyren, Cancellieri, and Triscone}}]{PhysRevLett.104.126803}
\bibinfo{author}{\bibfnamefont{A.~D.} \bibnamefont{Caviglia}},
  \bibinfo{author}{\bibfnamefont{M.}~\bibnamefont{Gabay}},
  \bibinfo{author}{\bibfnamefont{S.}~\bibnamefont{Gariglio}},
  \bibinfo{author}{\bibfnamefont{N.}~\bibnamefont{Reyren}},
  \bibinfo{author}{\bibfnamefont{C.}~\bibnamefont{Cancellieri}},
  \bibnamefont{and} \bibinfo{author}{\bibfnamefont{J.-M.}
  \bibnamefont{Triscone}}, \bibinfo{journal}{Phys. Rev. Lett.}
  \textbf{\bibinfo{volume}{104}}, \bibinfo{pages}{126803}
  (\bibinfo{year}{2010}),
  \urlprefix\url{https://link.aps.org/doi/10.1103/PhysRevLett.104.126803}.

\bibitem[{\citenamefont{Seiler et~al.}(2018)\citenamefont{Seiler, Zabaleta,
  Wanke, Mannhart, Kopp, and Braak}}]{PhysRevB.97.075136}
\bibinfo{author}{\bibfnamefont{P.}~\bibnamefont{Seiler}},
  \bibinfo{author}{\bibfnamefont{J.}~\bibnamefont{Zabaleta}},
  \bibinfo{author}{\bibfnamefont{R.}~\bibnamefont{Wanke}},
  \bibinfo{author}{\bibfnamefont{J.}~\bibnamefont{Mannhart}},
  \bibinfo{author}{\bibfnamefont{T.}~\bibnamefont{Kopp}}, \bibnamefont{and}
  \bibinfo{author}{\bibfnamefont{D.}~\bibnamefont{Braak}},
  \bibinfo{journal}{Phys. Rev. B} \textbf{\bibinfo{volume}{97}},
  \bibinfo{pages}{075136} (\bibinfo{year}{2018}),
  \urlprefix\url{https://link.aps.org/doi/10.1103/PhysRevB.97.075136}.

\bibitem[{\citenamefont{Baranger et~al.}(1993)\citenamefont{Baranger, Jalabert,
  and Stone}}]{Baranger1993}
\bibinfo{author}{\bibfnamefont{H.~U.} \bibnamefont{Baranger}},
  \bibinfo{author}{\bibfnamefont{R.~A.} \bibnamefont{Jalabert}},
  \bibnamefont{and} \bibinfo{author}{\bibfnamefont{A.~D.} \bibnamefont{Stone}},
  \bibinfo{journal}{Phys. Rev. Lett.} \textbf{\bibinfo{volume}{70}},
  \bibinfo{pages}{3876} (\bibinfo{year}{1993}),
  \urlprefix\url{https://link.aps.org/doi/10.1103/PhysRevLett.70.3876}.

\bibitem[{\citenamefont{Zozoulenko and Berggren}(1996)}]{Zozoulenko1996}
\bibinfo{author}{\bibfnamefont{I.~V.} \bibnamefont{Zozoulenko}}
  \bibnamefont{and} \bibinfo{author}{\bibfnamefont{K.-F.}
  \bibnamefont{Berggren}}, \bibinfo{journal}{Phys. Rev. B}
  \textbf{\bibinfo{volume}{54}}, \bibinfo{pages}{5823} (\bibinfo{year}{1996}),
  \urlprefix\url{https://link.aps.org/doi/10.1103/PhysRevB.54.5823}.

\bibitem[{\citenamefont{Cserti et~al.}(2004)\citenamefont{Cserti, Csord\'as,
  and Z\"ulicke}}]{Cserti2004}
\bibinfo{author}{\bibfnamefont{J.}~\bibnamefont{Cserti}},
  \bibinfo{author}{\bibfnamefont{A.}~\bibnamefont{Csord\'as}},
  \bibnamefont{and}
  \bibinfo{author}{\bibfnamefont{U.}~\bibnamefont{Z\"ulicke}},
  \bibinfo{journal}{Phys. Rev. B} \textbf{\bibinfo{volume}{70}},
  \bibinfo{pages}{233307} (\bibinfo{year}{2004}),
  \urlprefix\url{https://link.aps.org/doi/10.1103/PhysRevB.70.233307}.

\bibitem[{\citenamefont{Stornaiuolo et~al.}(2014)\citenamefont{Stornaiuolo,
  Gariglio, F\^ete, Gabay, Li, Massarotti, and Triscone}}]{PhysRevB.90.235426}
\bibinfo{author}{\bibfnamefont{D.}~\bibnamefont{Stornaiuolo}},
  \bibinfo{author}{\bibfnamefont{S.}~\bibnamefont{Gariglio}},
  \bibinfo{author}{\bibfnamefont{A.}~\bibnamefont{F\^ete}},
  \bibinfo{author}{\bibfnamefont{M.}~\bibnamefont{Gabay}},
  \bibinfo{author}{\bibfnamefont{D.}~\bibnamefont{Li}},
  \bibinfo{author}{\bibfnamefont{D.}~\bibnamefont{Massarotti}},
  \bibnamefont{and} \bibinfo{author}{\bibfnamefont{J.-M.}
  \bibnamefont{Triscone}}, \bibinfo{journal}{Phys. Rev. B}
  \textbf{\bibinfo{volume}{90}}, \bibinfo{pages}{235426}
  (\bibinfo{year}{2014}),
  \urlprefix\url{https://link.aps.org/doi/10.1103/PhysRevB.90.235426}.

\bibitem[{\citenamefont{Christensen et~al.}(2019)\citenamefont{Christensen,
  Trier, Niu, Gan, Zhang, Jespersen, Chen, and Pryds}}]{Rev2}
\bibinfo{author}{\bibfnamefont{D.~V.} \bibnamefont{Christensen}},
  \bibinfo{author}{\bibfnamefont{F.}~\bibnamefont{Trier}},
  \bibinfo{author}{\bibfnamefont{W.}~\bibnamefont{Niu}},
  \bibinfo{author}{\bibfnamefont{Y.}~\bibnamefont{Gan}},
  \bibinfo{author}{\bibfnamefont{Y.}~\bibnamefont{Zhang}},
  \bibinfo{author}{\bibfnamefont{T.~S.} \bibnamefont{Jespersen}},
  \bibinfo{author}{\bibfnamefont{Y.}~\bibnamefont{Chen}}, \bibnamefont{and}
  \bibinfo{author}{\bibfnamefont{N.}~\bibnamefont{Pryds}},
  \bibinfo{journal}{Advanced Materials Interfaces}
  \textbf{\bibinfo{volume}{6}}, \bibinfo{pages}{1900772}
  (\bibinfo{year}{2019}),
  \urlprefix\url{https://advanced.onlinelibrary.wiley.com/doi/abs/10.1002/admi.201900772}.

\bibitem[{\citenamefont{Groth et~al.}(2014)\citenamefont{Groth, Wimmer,
  Akhmerov, and Waintal}}]{Groth_2014}
\bibinfo{author}{\bibfnamefont{C.~W.} \bibnamefont{Groth}},
  \bibinfo{author}{\bibfnamefont{M.}~\bibnamefont{Wimmer}},
  \bibinfo{author}{\bibfnamefont{A.~R.} \bibnamefont{Akhmerov}},
  \bibnamefont{and} \bibinfo{author}{\bibfnamefont{X.}~\bibnamefont{Waintal}},
  \bibinfo{journal}{New Journal of Physics} \textbf{\bibinfo{volume}{16}},
  \bibinfo{pages}{063065} (\bibinfo{year}{2014}),
  \urlprefix\url{https://doi.org/10.1088/1367-2630/16/6/063065}.

\bibitem[{\citenamefont{Prawiroatmodjo
  et~al.}(2017)\citenamefont{Prawiroatmodjo, Leijnse, Trier, Christensen, Chen,
  Jespersen, Linderoth, and Pryds}}]{Prawiroatmodjo2017}
\bibinfo{author}{\bibfnamefont{G.}~\bibnamefont{Prawiroatmodjo}},
  \bibinfo{author}{\bibfnamefont{M.}~\bibnamefont{Leijnse}},
  \bibinfo{author}{\bibfnamefont{F.}~\bibnamefont{Trier}},
  \bibinfo{author}{\bibfnamefont{D.}~\bibnamefont{Christensen}},
  \bibinfo{author}{\bibfnamefont{Y.}~\bibnamefont{Chen}},
  \bibinfo{author}{\bibfnamefont{T.}~\bibnamefont{Jespersen}},
  \bibinfo{author}{\bibfnamefont{S.}~\bibnamefont{Linderoth}},
  \bibnamefont{and} \bibinfo{author}{\bibfnamefont{N.}~\bibnamefont{Pryds}},
  \bibinfo{journal}{Nature Communications} \textbf{\bibinfo{volume}{8}},
  \bibinfo{pages}{395} (\bibinfo{year}{2017}),
  \urlprefix\url{https://doi.org/10.1038/s41467-017-00495-7}.

\bibitem[{\citenamefont{Settino et~al.}(2021)\citenamefont{Settino, Citro,
  Romeo, Cataudella, and Perroni}}]{PhysRevB.103.235120}
\bibinfo{author}{\bibfnamefont{J.}~\bibnamefont{Settino}},
  \bibinfo{author}{\bibfnamefont{R.}~\bibnamefont{Citro}},
  \bibinfo{author}{\bibfnamefont{F.}~\bibnamefont{Romeo}},
  \bibinfo{author}{\bibfnamefont{V.}~\bibnamefont{Cataudella}},
  \bibnamefont{and} \bibinfo{author}{\bibfnamefont{C.~A.}
  \bibnamefont{Perroni}}, \bibinfo{journal}{Phys. Rev. B}
  \textbf{\bibinfo{volume}{103}}, \bibinfo{pages}{235120}
  (\bibinfo{year}{2021}),
  \urlprefix\url{https://link.aps.org/doi/10.1103/PhysRevB.103.235120}.

\bibitem[{\citenamefont{Jouan et~al.}(2020)\citenamefont{Jouan, Singh, Lesne,
  Vaz, Bibes, Barth\'el\'emy, Ulysse, Stornaiuolo, Salluzzo, Hurand
  et~al.}}]{Jouan2020}
\bibinfo{author}{\bibfnamefont{A.}~\bibnamefont{Jouan}},
  \bibinfo{author}{\bibfnamefont{G.}~\bibnamefont{Singh}},
  \bibinfo{author}{\bibfnamefont{E.}~\bibnamefont{Lesne}},
  \bibinfo{author}{\bibfnamefont{D.~C.} \bibnamefont{Vaz}},
  \bibinfo{author}{\bibfnamefont{M.}~\bibnamefont{Bibes}},
  \bibinfo{author}{\bibfnamefont{A.}~\bibnamefont{Barth\'el\'emy}},
  \bibinfo{author}{\bibfnamefont{C.}~\bibnamefont{Ulysse}},
  \bibinfo{author}{\bibfnamefont{D.}~\bibnamefont{Stornaiuolo}},
  \bibinfo{author}{\bibfnamefont{M.}~\bibnamefont{Salluzzo}},
  \bibinfo{author}{\bibfnamefont{S.}~\bibnamefont{Hurand}},
  \bibnamefont{et~al.}, \bibinfo{journal}{Nature Electronics}
  \textbf{\bibinfo{volume}{3}}, \bibinfo{pages}{201} (\bibinfo{year}{2020}),
  \bibinfo{note}{published: 16 March 2020},
  \urlprefix\url{https://doi.org/10.1038/s41928-020-0383-2}.

\bibitem[{\citenamefont{Vicente-Arche et~al.}(2021)\citenamefont{Vicente-Arche,
  Bréhin, Varotto, Cosset-Cheneau, Mallik, Salazar, Noël, Vaz, Trier,
  Bhattacharya et~al.}}]{Arche}
\bibinfo{author}{\bibfnamefont{L.~M.} \bibnamefont{Vicente-Arche}},
  \bibinfo{author}{\bibfnamefont{J.}~\bibnamefont{Bréhin}},
  \bibinfo{author}{\bibfnamefont{S.}~\bibnamefont{Varotto}},
  \bibinfo{author}{\bibfnamefont{M.}~\bibnamefont{Cosset-Cheneau}},
  \bibinfo{author}{\bibfnamefont{S.}~\bibnamefont{Mallik}},
  \bibinfo{author}{\bibfnamefont{R.}~\bibnamefont{Salazar}},
  \bibinfo{author}{\bibfnamefont{P.}~\bibnamefont{Noël}},
  \bibinfo{author}{\bibfnamefont{D.~C.} \bibnamefont{Vaz}},
  \bibinfo{author}{\bibfnamefont{F.}~\bibnamefont{Trier}},
  \bibinfo{author}{\bibfnamefont{S.}~\bibnamefont{Bhattacharya}},
  \bibnamefont{et~al.}, \bibinfo{journal}{Advanced Materials}
  \textbf{\bibinfo{volume}{33}}, \bibinfo{pages}{2102102}
  (\bibinfo{year}{2021}),
  \urlprefix\url{https://advanced.onlinelibrary.wiley.com/doi/abs/10.1002/adma.202102102}.

\bibitem[{Note1()}]{Note1}
Note1, \bibinfo{note}{in our simulations, the leads are modeled as
  semi-infinite electrodes, translationally invariant along the
  \(y\)-direction, with the transverse motion (along \(x\)) quantized into five
  propagating modes. Each mode has its own Fermi wavevector \(k_F\), and for
  the five modes we find \(k_Fa \in [0.7, 1.8]\), which corresponds to
  \(\lambda _F/a \in [3.48, 8.97]\)}.

\bibitem[{\citenamefont{Beenakker}(1997)}]{RevModPhys.69.731}
\bibinfo{author}{\bibfnamefont{C.~W.~J.} \bibnamefont{Beenakker}},
  \bibinfo{journal}{Rev. Mod. Phys.} \textbf{\bibinfo{volume}{69}},
  \bibinfo{pages}{731} (\bibinfo{year}{1997}),
  \urlprefix\url{https://link.aps.org/doi/10.1103/RevModPhys.69.731}.

\bibitem[{\citenamefont{Jovanovi\ifmmode~\acute{c}\else \'{c}\fi{}
  et~al.}(2010)\citenamefont{Jovanovi\ifmmode~\acute{c}\else \'{c}\fi{},
  Fruchter, Li, and Raffy}}]{PhysRevB.81.134520}
\bibinfo{author}{\bibfnamefont{V.~P.}
  \bibnamefont{Jovanovi\ifmmode~\acute{c}\else \'{c}\fi{}}},
  \bibinfo{author}{\bibfnamefont{L.}~\bibnamefont{Fruchter}},
  \bibinfo{author}{\bibfnamefont{Z.~Z.} \bibnamefont{Li}}, \bibnamefont{and}
  \bibinfo{author}{\bibfnamefont{H.}~\bibnamefont{Raffy}},
  \bibinfo{journal}{Phys. Rev. B} \textbf{\bibinfo{volume}{81}},
  \bibinfo{pages}{134520} (\bibinfo{year}{2010}),
  \urlprefix\url{https://link.aps.org/doi/10.1103/PhysRevB.81.134520}.

\bibitem[{\citenamefont{Jovanovi\ifmmode~\acute{c}\else \'{c}\fi{}
  et~al.}(2021)\citenamefont{Jovanovi\ifmmode~\acute{c}\else \'{c}\fi{}, Raffy,
  Li, Rem\'enyi, and Monceau}}]{PhysRevB.103.014520}
\bibinfo{author}{\bibfnamefont{V.~P.}
  \bibnamefont{Jovanovi\ifmmode~\acute{c}\else \'{c}\fi{}}},
  \bibinfo{author}{\bibfnamefont{H.}~\bibnamefont{Raffy}},
  \bibinfo{author}{\bibfnamefont{Z.~Z.} \bibnamefont{Li}},
  \bibinfo{author}{\bibfnamefont{G.}~\bibnamefont{Rem\'enyi}},
  \bibnamefont{and} \bibinfo{author}{\bibfnamefont{P.}~\bibnamefont{Monceau}},
  \bibinfo{journal}{Phys. Rev. B} \textbf{\bibinfo{volume}{103}},
  \bibinfo{pages}{014520} (\bibinfo{year}{2021}),
  \urlprefix\url{https://link.aps.org/doi/10.1103/PhysRevB.103.014520}.

\bibitem[{\citenamefont{Prentice and Coldea}(2016)}]{PhysRevB.93.245105}
\bibinfo{author}{\bibfnamefont{J.~C.~A.} \bibnamefont{Prentice}}
  \bibnamefont{and} \bibinfo{author}{\bibfnamefont{A.~I.}
  \bibnamefont{Coldea}}, \bibinfo{journal}{Phys. Rev. B}
  \textbf{\bibinfo{volume}{93}}, \bibinfo{pages}{245105}
  (\bibinfo{year}{2016}),
  \urlprefix\url{https://link.aps.org/doi/10.1103/PhysRevB.93.245105}.

\end{thebibliography}
\end{document}